\documentclass[review]{elsarticle}
\usepackage[commentmarkup=todo,authormarkup=none]{changes}
\biboptions{sort&compress,numbers}
\usepackage[margin=1in]{geometry}
\usepackage{float}
\usepackage[colorlinks=false]{hyperref}\hypersetup{pdfborder={0 0 0}}
\usepackage{miller}
\usepackage{bm,amsmath,amssymb,amsthm}
\usepackage{textcomp, gensymb}
\usepackage{enumitem}
\usepackage{mathalfa}
\usepackage{cancel}
\usepackage{subfigure}\usepackage{graphicx}
\DeclareUnicodeCharacter{03B3}{$\gamma$}
\DeclareUnicodeCharacter{03A3}{$\Sigma$}
\DeclareUnicodeCharacter{2009}{ }
\usepackage[T1]{fontenc}
\usepackage[margin=1in]{geometry}
\usepackage{booktabs,dcolumn,caption}
\newcolumntype{d}[1]{D{.}{.}{#1}} % "decimal" column type
\renewcommand{\ast}{{}^{\textstyle *}} % for raised "asterisks"

\DeclareUnicodeCharacter{3B1}{\ensuremath{\alpha}}
\DeclareUnicodeCharacter{03C3}{\ensuremath{\gamma}}
\newcommand{\rmd}{\mathrm{d}}

\definecolor{C0}{HTML}{1F77B4}
\definecolor{C1}{HTML}{FF7F0E}
\definechangesauthor[color=C0]{R1} % changes relevant to R1
\definechangesauthor[color=C1]{R2} % changes relevant to R2

\begin{document}
\begin{frontmatter}
\title{Kinetic and Equilibrium Shapes of Cylindrical Grain Boundaries }
\author[cmu]{Anqi Qiu}
\ead{anqiq@andrew.cmu.edu}

\author[cityu]{Caihao Qiu}
\ead{caihaoqiu2-c@my.cityu.edu.hk}

\author[lanl]{Ian Chesser}
\ead{ichesser@lanl.gov}

\author[cityu]{Jian Han}
\ead{jianhan@cityu.edu.hk}

\author[hku]{David J. Srolovitz}
\ead{srol@hku.hk}

\author[cmu,um]{Elizabeth A. Holm}
\ead{eaholm@umich.edu}

\address[cmu]{Department of Materials Science and Engineering, Carnegie Mellon University, Pittsburgh, PA 15213, USA}

\address[cityu]{Department of Materials Science and Engineering, City University of Hong Kong, Hong Kong Special Administrative Region, China}

\address[lanl]{XCP-5, Los Alamos National Laboratory, Los Alamos, NM 87545}

\address[hku]{Department of Mechanical Engineering, The University of Hong Kong, Hong Kong Special Administrative Region, China}

\address[um]{Department of Materials Science and Engineering, University of Michigan, Ann Arbor, MI 48109}

\begin{abstract}
In this work, we investigate the shape evolution of rotated, embedded, initially cylindrical grains (with [001] cylinder axis) in Ni under an applied synthetic driving force via molecular dynamics simulations and a continuum, disconnection-based grain boundary migration model. For some initial misorientations, the expanding grains form  well-defined, faceted shapes, while for others the shapes remain cylindrical. The embedded grains tend to rotate during grain boundary migration, with the direction of rotation dependent on initial misorientation and direction of growth (expand/shrink). The kinetic shapes, which are bounded by low mobility grain boundary planes, differ from equilibrium shapes (bounded by low energy grain boundaries). The multi-mode disconnection model-based predictions are consistent with the molecular dynamics results for faceting tendency, kinetic grain shape,  and grain rotation as a function of misorientation and whether the grains are expanding or contracting. 
This demonstrates that grain boundary migration and associated grain rotation are mediated by disconnection flow along grain boundaries.
\end{abstract}
  
\end{frontmatter}

%\linenumbers


\section{Introduction}

Facet formation on grain boundaries during grain growth is a widely observed phenomenon in polycrystalline materials. Considered as the manifestation of grain boundary energy anisotropy, faceting in grain boundaries is often discussed in terms of energy. Grain boundary energy is highly anisotropic with respect to misorientation and boundary plane inclination \cite{rohrer_grain_2011}. In most polycrystalline materials, the distribution of grain boundary planes is strongly related to grain boundary energy anisotropy \cite{saylor_habits_2004}. High energy grain boundaries have been shown to form low energy facets to reduce system energy \cite{goodhew_low_1978,muschik_energetic_1993,barg_faceting_1995,dobrovolski2024facet}. The flat facets often form sawtooth profiles consisting of facets with the same misorientation and different boundary plane inclinations \cite{merkle_low-energy_1992}. During grain growth, higher energy grain boundaries are replaced by lower energy grain boundaries, leading to a decrease in overall energy per volume \cite{xu_energy_2023,xu_grain_2024}.

The faceting behavior at interfaces is often investigated by minimizing the interfacial energy. For example, the equilibrium shape of a crystal suspended in melt can be obtained from minimizing the interfacial energy under constant volume, through a process known as Wulff construction \cite{wulff_xxv_1901, herring_theorems_1951}. Bound by lowest energy planes, the equilibrium crystal shape, also known as the Wulff shape, can be obtained from a polar plot of interfacial energy, as the inner envelope of perpendicular planes to the radial interfacial energy vectors. Similarly, the Wulff shape of an isolated grain embedded in a matrix grain can also be constructed using a polar plot of grain boundary energy. In many cases, the facets formed by grain boundaries do correspond to facets in the computed Wulff shapes \cite{goodhew_low_1978,muschik_energetic_1993,barg_faceting_1995,merkle_low-energy_1992}. 

Though some faceting phenomena can indeed be explained by grain boundary energy anisotropy, it is not the only factor that influences faceting. In crystal growth under non-equilibrium conditions, the crystal shapes do not form facets according to surface energy anisotropy, but instead evolve toward their "kinetic Wulff shapes", bound by the slowest moving planes \cite{sekerka_equilibrium_2005}. Similarly, under non-equilibrium conditions, the shapes of grains can also evolve toward their kinetic shapes, bound by the slowest moving grain boundary planes. Rheinheimer \textit{et al.} discovered that in polycrystalline microstructures, boundaries are dominated by the planes in the kinetic shapes at the early stages of grain growth, and by the planes in the Wulff shapes during later stages \cite{rheinheimer_equilibrium_2020}. 

In addition to energy, stress can also be a driving force for facet formation \cite{hanyu_stress-induced_2005}. The motions of interfaces in crystalline solids, including grain boundaries, are mediated by disconnections, which are line defects with step and/or dislocation characters. The elastic fields of the grain boundaries, associated with the dislocation character of the grain boundary dislocations and disconnections, affect the migration and faceting of grain boundaries. Some of the present authors formulated a continuum disconnection-based thermodynamic and kinetic model incorporating bicrystallography, interfacial structure, interfacial energy, and elasticity effects to investigate the shape evolution of embedded cylindrical grains, which demonstrates that the elastic effects of disconnections enhance faceting and modify grain boundary facet morphologies \cite{qiu_interface_2023}. Other work by the same authors showed that grain rotation may be induced by the flow of disconnections along grain boundaries \cite{qiu_disconnection_2024}.

In this work, we investigate the kinetic shapes of cylindrical grains embedded in a matrix grain through molecular dynamics (MD) simulations. In the MD simulations, a Synthetic Driving Force (SDF) method, which has been previously used to investigate the quasi-equilibrium shapes of embedded spherical grains \cite{schratt_grain_2021}, is implemented to enable the expansion of the cylindrical grains at elevated temperatures. The observed kinetic shapes from MD simulations are then compared with numerical simulations using continuum disconnection models to reveal the mechanisms of facet growth and associated grain rotations. The Wulff shapes of the investigated boundaries at 0 K are computed using molecular statics (MS) simulations. 

\section{Methods}
\label{sec:sample2}

\subsection{Molecular dynamics simulations}
All the molecular dynamics (MD) simulations are performed using the Large-scale Atomic/Molecular Massively Parallel Simulator (LAMMPS) software package \cite{plimpton_fast_1995,thompson_lammps_2022}. Ni, a typical face-centered cubic (FCC) metallic material widely used in the study of grain boundary motion, is used as the model material. Atomic interactions in the system are modelled with the Foiles-Hoyt EAM interatomic potential for Ni \cite{foiles_computation_2006}. The melting point associated with the potential is 1565 K.

Initially, the simulation box is constructed by creating a block containing perfect crystal structure, with edges aligned with <100> crystal directions. Next, a new cylindrical grain is created by rotating a cylindrical portion by the [001] direction (z-axis) by a certain misorientation angle $\theta$. After the cylindrical grain is created, pairs of atoms whose separation distances are within the specific cutoff distance of 1.6 Å of each other are searched for, and one of them is deleted. The constructed grain boundary structure is a cylindrical [001] tilt boundary, with misorientation angle of $\theta$, and a full circle of boundary plane inclinations $0-360 \degree$. The simulation box is created with periodic boundary conditions in all three Cartesian directions. The grain boundary structure possesses ${D_{4h}}$ point group symmetry, so only $\theta = 0-45 \degree$ is needed to explore the entire misorientation range.

Before the structure is heated, molecular statics (MS) simulations are performed using the Polak-Ribiere version of the conjugate descent algorithm \cite{hestenes_methods_1952} at 0 K to minimize the system energy. After energy minimization, the system is heated by assigning random initial atomic velocities sampled from the Maxwell-Boltzmann distribution according to the desired system temperature. The evolution of the system is then investigated by MD simulations. The isothermal-isobaric (NPT) ensemble is implemented to ensure that as the grain boundary shrinks during migration, the changes in grain boundary free volume will not cause significant internal stresses \cite{trautt_grain_2012}. In the simulations, the pressure is set to zero in all three Cartesian directions, and the temperature is set to a finite value between 700-1400 K, and the timestep is 0.003 ps. Unless specified otherwise, the simulation temperature is set to 1000 K. In the LAMMPS simulations, different initial velocity distributions at a given temperature are specified with different initial velocity seeds. In a previous study, some of the present authors demonstrated that the initial velocity distributions of atoms can significantly impact grain boundary migration, when the cylindrical grains are left to shrink with no externally applied driving force \cite{qiu_variability_2023}. Therefore, multiple parallel simulation runs differing only in initial velocity seeds may be needed for the same initial structure.

The simulation setup is similar to that of the shrinking cylindrical grain simulations described in a previous publication by some of the present authors \cite{qiu_variability_2023}. The only differences are in the addition of a Synthetic Driving Force (SDF) to enable the expansion of the cylindrical grain, and adjustments to system size to accommodate the growing grain. The initial radius of the inner grain is set to 15${a_0}$ (53.8Å), the margin is set to 30${a_0}$ (105.6Å) or larger, and the box thickness along the tilt axis is set to 5${a_0}$ (17.6Å).

In most of our simulations, we do not apply any additional constraints to the system. With no additional constraints, the embedded cylindrical grains may rotate, as the SDF does not conserve linear or angular momentum. Grain rotation can be prohibited by keeping a small group of atoms (a smaller cylinder with radius of 5$a_0$) in the center of the cylindrical grain immobile, though there will still be some lattice distortions related to the application of the SDF. The grain boundaries with a small group of fixed atoms to prevent grain rotations are termed "fixed boundaries", and the boundaries with no fixed atoms are termed "unfixed boundaries". The fixed boundary simulations are performed to emulate the grain growth process in realistic materials, where grain rotations are not typically observed. 

\subsection{Synthetic driving force}

In order to facilitate the expansion of the inner cylindrical grain by counteracting the intrinsic curvature driving force, we employ a Synthetic Driving Force (SDF) method, first proposed by Janssens \cite{janssens_computing_2006}. This method introduces an artificial energy difference between the two grains by adding a potential energy to each atom according its crystal orientation. 

For this study, we implement a more efficient energy-conserving orientational (ECO) driving force method \cite{schratt_efficient_2020}, as proposed by Schratt and Mohles. Initially, a symmetric orientational order parameter for atom $i$ is defined as:
\begin{equation}\label{order}
\chi_{i} = \frac{1}{N}\sum_{k=1}^{3}[|\psi_k^I(\mathbf{r_i})|^2-|\psi_k^{II}(\mathbf{r_i})|^2]
\end{equation}
where N indicates the normalization factor at $T = 0 K$, and the complex functions:
$\psi_k^I(\mathbf{r_i})$ and $\psi_k^{II}(\mathbf{r_i})$
evaluate how closely an atom $i$ matches the perfect orientations of crystals I and II:
\begin{equation}\label{complex}
\psi_k^{\lambda}(\mathbf{r_i}) = \sum_{j \in \Gamma_i} w_{ij} \exp(i(\mathbf{r_i} - \mathbf{r_j}) \mathbf{q_k^{\lambda}})
\end{equation}
where $\lambda =$ I, II,  $\mathbf{q_k^{\lambda}}  (k=1,2,3)$ are independent reciprocal lattice vectors, $j \in \Gamma_i$ indicates sum over neighbor atoms, and $w_{ij} = w(|\mathbf{r_i} - \mathbf{r_j}|)$ is the weight function. Inside a sphere centered at the atomic position $\mathbf{r_i}$, with radius of $r_c$, the weight function smoothly transitions from $w(0) = 1$ to $w(r_c) = 0$. The weight function is zero outside the cutoff radius of $r_c$, so that only nearest neighbors are included in the calculation. 

The order parameter $\chi_i$ is normalized to the range of [-1,+1] at $T = 0 K$. Its value can exceed the range at higher temperatures, so a cutoff parameter $\eta$ is used to truncate the thermal fluctuations inside bulk grains, in this study, $\eta = 0.25$. 

A new order parameter $X$ is defined as:
\begin{equation}\label{X}
X_{i} = \begin{cases}
+1, & \chi_{i} \geq \eta\\
\sin{\frac{\pi \chi_{i}}{2 \eta}}, & -\eta < \chi_{i} < \eta\\ -1, & \chi_{i} \leq -\eta
\end{cases} 
\end{equation}

The artificial energy can be added to each atom according to the order parameter:
\begin{equation}\label{u}
u(\chi_{i}) = \frac{u_0}{2} X_{i} = \frac{u_0}{2} \begin{cases}
+1, & \chi_{i} \geq \eta\\
\sin{\frac{\pi \chi_{i}}{2 \eta}}, & -\eta < \chi_{i} < \eta\\ -1, & \chi_{i} \leq -\eta
\end{cases} 
\end{equation}

\subsection{Grain segmentation}
\begin{figure}[ht!]
    \centering\leavevmode
    \includegraphics[width=0.9\textwidth]{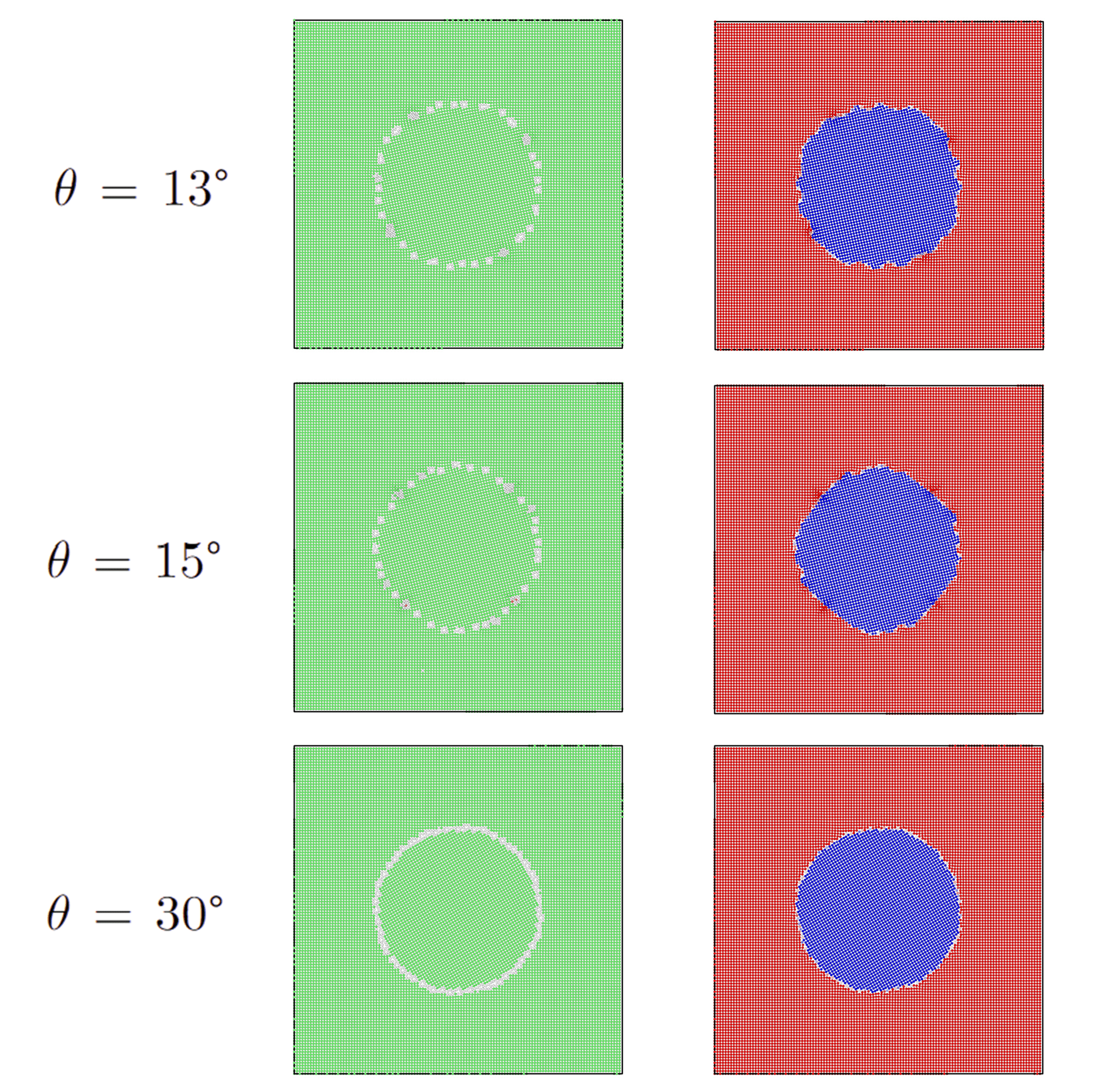}
    \caption{Embedded grains with initial misorientations $13 \degree$, $15 \degree$, and $30 \degree$ at 0 K, colored by polyhedral template matching (left) and orientation parameter (right).}
    \label{fig:13-15-30}
\end{figure}

The orientation-dependent order parameter of the ECO driving force, as defined by Schratt and Mohles \cite{schratt_efficient_2020}, is used for grain segmentation. In order to evaluate the performance of the order parameter in segmenting grains, the polyhedral template matching method (PTM) \cite{larsen_robust_2016}, as implemented in OVITO \cite{stukowski_visualization_2009}, is also employed to identify grain boundary atoms. The embedded grains with initial misorientations of $13 \degree$, $15 \degree$, and $30 \degree$ at 0 K, colored by PTM and order parameter, are shown in Figure \ref{fig:13-15-30}. In systems segmented by the PTM algorithm, atoms in the perfect lattice are colored green and atoms in the grain boundary are non-green (left column of Figure \ref{fig:13-15-30}). In the grain boundary surrounding the grain with $\theta = 30 \degree$ misorientation, referred to here as the $\theta = 30 \degree$ grain boundary, for example, the identified grain boundary structure is composed of a circle of distinct kite-shaped structural units very close to each other. In systems segmented by the order parameter, the atoms with order parameter $X_{i} = -1$, colored in red, are identified as the outer grain, and atoms with $X_{i} = 1$, colored blue, are identified as the inner cylindrical grain (right column of Figure \ref{fig:13-15-30}). Grain boundary atoms are identified as those with order parameter $-1 < X_{i} < 1$. Both PTM and order parameter identify the grain boundary as a circle of atoms around the interface between the two grains, therefore, the order parameter is very successful in segmenting two grains separated by the $\theta = 30 \degree$ boundary at 0K.

The $\theta = 13 \degree$ and $\theta = 15 \degree$ boundaries, which are both considered low angle boundaries, are composed of distinct kite-shaped structural units spaced apart from each other. Due to the low misorientation, it is challenging to distinguish between the two grains, especially in the regions surrounding the grain boundary. For the $\theta = 13 \degree$ boundary, the inner grain segmented by the order parameter is not cylindrical. Instead, the boundary appears corrugated, due to ambiguities in grain membership.

When the misorientation is sufficiently high ($\theta >= 15 \degree$), the order parameter can be used successfully for segmenting grains. When the misorientation is too low ($\theta < 15 \degree$), the order parameter can no longer be used effectively for grain segmentation. 

\clearpage
\subsection{Wulff shape construction}

The faceted shape of an embedded grain is often investigated by minimizing the interfacial energy through Wulff construction. The geometric method of obtaining the equilibrium shape of a crystal suspended in melt from the $\gamma$-plot of the orientation dependence of interfacial energy under constant volume was first proposed by Wulff without mathematical proof in 1901 \cite{wulff_xxv_1901}. The equilibrium conditions for crystal growth was later extended and proved mathematically by Herring \cite{herring_theorems_1951}. Known as the Wulff-Herring construction, the method is used to determine the Wulff shapes by drawing perpendicular planes at each point in the $\gamma$-plot and taking the enveloped shape. The facets correspond to cusps in the $\gamma$-plot. For crystals suspended in melt, the constructed Wulff shapes are the equilibrium shapes of the crystals. The Wulff-Herring construction has also been extended to grain boundaries \cite{goodhew_low_1978,muschik_energetic_1993,barg_faceting_1995,merkle_low-energy_1992,straumal_faceting_2001,straumal_temperature_2006}, and is a useful tool for the prediction of faceting. The equilibrium shape of an embedded grain with fixed volume can also be found through Wulff construction. 

In order to compute the Wulff shape of a cylindrical [001] tilt grain boundary, the grain boundary energies at boundary plane inclination angles $0-360 \degree$ are required to construct the $\gamma$-plot. Therefore, there is a need to accurately calculate the energies of grain boundaries with arbitrary misorientation and boundary plane inclination angles that do not belong to any coincident-site lattice (CSL).

The grain boundary energies of arbitrary planar grain boundaries can be computed using a cutoff sphere molecular dynamics model \cite{li_atomistic_2019}. This method builds upon Lee and Choi's method of constructing spherical simulation cells with free surfaces \cite{lee_computation_2004}. In order to reduce the effect of the interaction between the surface and the grain boundary, the cutoff sphere model uses only an internal portion of the spherical simulation cell to calculation grain boundary energy. 

Lee and Choi's method is summarized as follows. To construct the spherical simulation cell containing an arbitrary grain boundary, a monocrystalline sphere with radius $r$ is first constructed. The potential energy of the sphere $E_1$ is calculated by MS simulation using a conjugate gradient energy minimization \cite{hestenes_methods_1952}. The monocrystalline sphere is then rotated to form two additional spheres with different crystalline orientations $\mathbf{g_A}$ and $\mathbf{g_B}$, with the desire misorientation $\Delta \mathbf{g} = \mathbf{g_A}^{-1} \mathbf{g_B}$ between them. At the desired boundary plane inclination, which has a boundary plane normal $\mathbf{n}$, the two spheres are cut into hemispheres, and joined together to form a new sphere containing the desired grain boundary structure. The potential energy of the sphere containing the grain boundary $E_2$ is also calculated through MS simulation. The area of the grain boundary $A = \pi r^2$. The grain boundary energy $E_{gb}$ is then estimated through the energy difference between the monocrystalline sphere and the sphere containing the grain boundary:

\begin{equation}\label{gbs}
E_{gb} = \frac{E_2-E_1}{A} = \frac{E_2-E_1}{\pi r^2}
\end{equation}

\begin{figure}[ht!]
    \centering\leavevmode
    \includegraphics[width=1\textwidth]{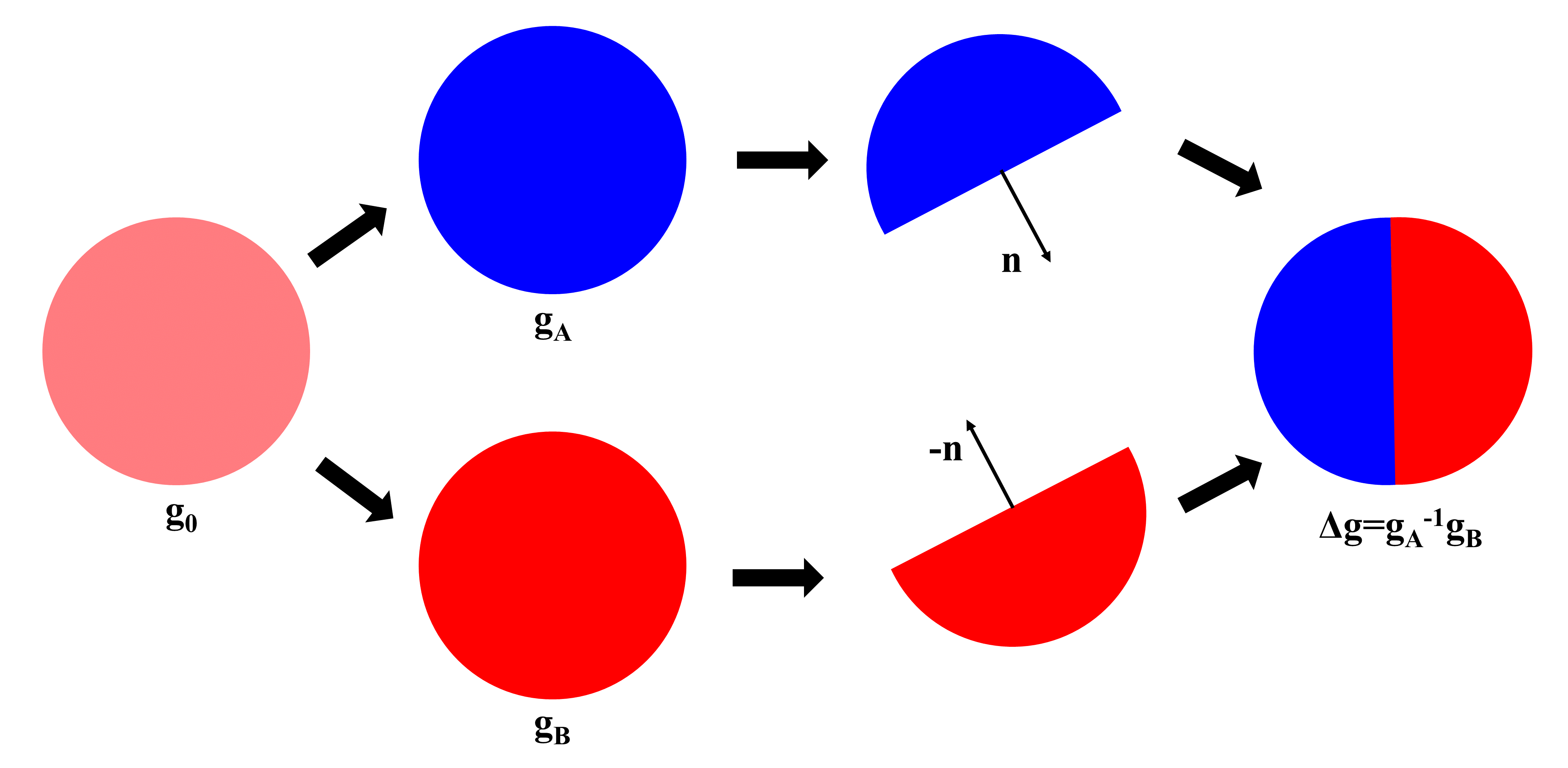}
    \caption{Construction of a spherical simulation cell containing an arbitrary grain boundary, replicated from \cite{li_atomistic_2019}.}
    \label{fig:sphere}
\end{figure} 

Figure \ref{fig:sphere} shows the construction process of a spherical simulation cell to compute the grain boundary energy of arbitrary grain boundaries. This method is based on the assumption that the surface energies of the two spheres are the same, which is not true because the energies of the atoms at the regions where the grain boundary interacts with the surface are different from the energies of other surface atoms. In order to eliminate the effect of the interactions between the surface and the grain boundary, Li \textit{et al.} developed a cutoff sphere model, which only uses an inner portion of the spherical simulation cell without the surface atoms to compute grain boundary energy \cite{li_atomistic_2019}. The energy $E_c$ of the inner sphere with radius $r_c$ is calculated through MS simulation. For computing the minimum energy of $E_c$, the deletion of over-close atoms in the grain boundary region is needed. Pairs of atoms whose separation distances are within a critical distance $d_c$ of each other are searched for, and one of them is deleted. The critical distance value ranges from 0.005$a_0$ to 0.700$a_0$, with intervals of 0.005$a_0$. The critical distance at which $E_c$ is the minimum is selected, and the number of atoms in the inner sphere is recorded as $N_c$. From a separate simulation using a perfect crystalline block, the cohesive energy $E_{coh}$ is calculated. The grain boundary area $A_c = \pi r_c^2$. The grain boundary energy $E_{gb}$ can be calculated as:

\begin{equation}\label{gbcs}
E_{gb} = \frac{E_c-N_c E_{coh}}{A_c} = \frac{E_c-N_c E_{coh}}{\pi r_c^2}
\end{equation}

In this work, the radius of the spherical simulation cell $r = 30 a_0$ (105.6Å), and the radius of the inner sphere $r_c = 15 a_0$ (52.8Å). The large distance between the surface of the inner sphere and the surface of the outer sphere ensures that the portion of grain boundary atoms in the inner sphere have minimal interactions with the surface. The size of the inner sphere is sufficiently large so that the calculated grain boundary energy is not size-dependent. However, limitations of the calculation include the sampling of only one patch of an non-periodic interface and the exploration of a limited range of atomic densities in the grain boundary region.

\begin{figure}[ht!]
    \centering\leavevmode
    \includegraphics[width=1\textwidth]{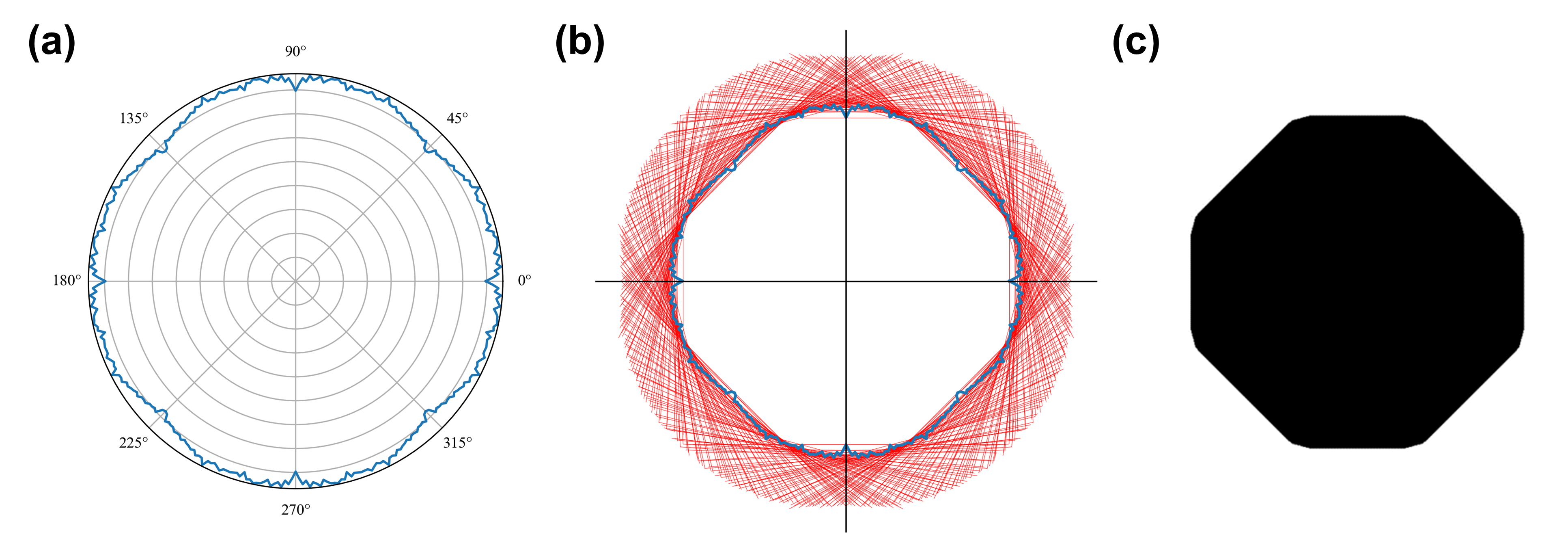}
    \caption{(a) Polar plot of the $\theta = 30 \degree$ boundary (b) Wulff shape construction using the polar plot (c) Constructed Wulff shape.}
    \label{fig:Wulff-construction}
\end{figure}

For each misorientation, since the cylindrical grain boundary possesses ${D_{4h}}$ point group symmetry, only the inclination angles $0-45 \degree$, in $1 \degree$ intervals, are computed to obtain the full $0-360 \degree$ range. The calculated grain boundary energies for different inclination angles are plotted radially in a polar plot. The Wulff shape construction consists of three steps. First, radial vectors are drawn from the origin to each point on the polar plot. Second, perpendicular planes are drawn at each point the radial vector intersects the polar plot. Finally, the shape enveloped by all the perpendicular planes is determined as the Wulff shape. Figure \ref{fig:Wulff-construction} shows the Wulff shape construction process for the $\theta = 30 \degree$ boundary. The shape enveloped by all the perpendicular planes is octagonal.

\subsection{Continuum disconnection models}

We recently proposed a bicrystallography-respecting continuum model to describe the disconnection-mediated migration of an arbitrarily-curved grain boundary \cite{han2022disconnection}.
Disconnections are the line defects constrained on the grain boundaries (and other general interfaces) possessing both step and/or dislocation character and, thus, can be characterized by their step heights $h$ and Burgers vectors ($\mathbf{b}$); the shear-coupling factor of the grain boundary migration can be expressed as $\beta = |\mathbf{b}|/ h$.
An arbitrarily-curved grain boundary can be considered as the combination of disconnections of corresponding reference interfaces, i.e., crystallographic planes with a high density of coincidence sites in a CSL lattice.
The equation of motion for an arbitrarily curved grain boundary can be expressed as \cite{han2022disconnection}
\begin{equation}\label{single_EOM}
\mathbf{v} = \mathbf{M} (\Gamma \kappa + \boldsymbol{\tau}\cdot\boldsymbol{\beta} + \psi) \hat{\mathbf{n}},
\end{equation}
where $\mathbf{v}$ is the grain boundary migration velocity, $\mathbf{M}$ is the grain boundary mobility tensor. 
$\Gamma \kappa$ and $\psi$ represents the capillarity force and synthetic driving force acting on the step character of disconnections, where $\Gamma$ is the grain boundary stiffness, $\kappa$ is the local grain boundary curvature and $\psi$ is the chemical free energy density jump across the interface.
$\boldsymbol{\tau}\cdot\boldsymbol{\beta}$ represents the elastic driving force acting on the dislocation character of disconnections, where $\boldsymbol{\tau}$ is the resolved shear stress and $\boldsymbol{\beta}$ is the shear-coupling factor vectors of the reference interfaces (i.e., highly-coherent low-energy planes in the CSL lattice of this arbitrarily-curved grain boundary.).
We note that the elastic driving force can be caused by both self-elastic interactions among the disconnection Burgers vector and the external stress, see the details in literature \cite{qiu_interface_2023}.

The same grain boundary can activate different disconnections with different step heights ($h_m$) and Burgers vectors ($b_m$) during its migration (i.e., different disconnection modes, $m$ represents the mode) \cite{chen_grain_2019, han_grain-boundary_2018}. 
At low temperature, only the disconnection mode with lowest activation energy will dominate the grain boundary migration (described by Equation \eqref{single_EOM}), whilst more disconnection modes will be involved at high temperature.

We also extend the equation of motion to take the multiple disconnection modes into consideration, which can be expressed as \cite{qiu_disconnection_2024}
\begin{equation}\label{single_EOM1}
\mathbf{v} = \mathbf{M}_x \sum_m(\Gamma \hat{\mathbf{l}}_m^\prime +  \tau\boldsymbol{\Lambda}_m + \psi \hat{\mathbf{n}}),
\end{equation}
where $\hat{\mathbf{l}}_m^\prime$ represents the local curvature expressed by the first derivative of the density of the disconnection mode $m$, $\boldsymbol{\Lambda}_m$ refers to the term with shear-coupling factor of disconnection mode $m$.
Note that, when the boundary migration is mediated by single disconnection mode, the disconnection densities are bijectively related to the shape of the grain boundary.
This bijection will be broken when more than one disconnection modes are involved.
Hence, different from the single disconnection-mode case, an additional equation of motion is required to describe the evolution of the densities of different disconnection modes.
The details of the derivation and explanation of the multiple disconnection mode-mediated grain boundary migration can be found in Section 2 of the Supplementary Materials and the literature \cite{qiu_disconnection_2024}.

We conduct a series of numerical simulations to study the faceting behavior and grain rotation of the embedded grains with three typical misorientation angles, i.e., $\Sigma17[100]$ ($\theta = 28.1\degree$), $\Sigma5[100]$ ($\theta = 36.9\degree$), $\Sigma29[100]$ ($\theta = 43.6\degree$), which are equivalent to [001] boundaries with the same $\Sigma$ values.
Numerical simulations are performed by solving the equations of motion mentioned above and in Supplementary Materials by a finite difference method using a three-point stencil. 
We employ reduced quantities in the numerical simulations (see \cite{han2022disconnection, qiu_disconnection_2024} for details). 
The shape of the initial embedded grain in all simulations is a circle of radius 100 (see the red dash lines in Figures \ref{fig:shape-continuum}(d)-(i)). 
The parameters (i.e., those for grain boundary bicrystallography, thermodynamics, and kinetics) used for the numerical simulations are given in Supplementary Tables 1 and 2 of the Supplementary Materials. 
The rotation field is calculated as the skew part of the displacement gradient \cite{zhang2014lattice, kareer2020scratching}.

\section{Results and Discussions}
\label{sec:results}
\subsection{Kinetic shapes under synthetic driving force }

We apply driving forces of different magnitudes to enable the growth of the inner cylindrical grains, and observe the evolution of the surrounding grain boundaries. The grains have initial misorientations of $\theta = 15-45 \degree$, and no additional constraints are applied, and the grain boundaries are referred to as the $\theta = 15-45\degree$ unfixed boundaries here. The exact magnitude of the driving force is insignificant to the shape evolution process. As long as the applied driving force is sufficient to enable grain growth, but not so large that the system is broken apart, the observed shapes are the same.  

\begin{figure}[ht!]
    \centering\leavevmode
    \includegraphics[width=1\textwidth]{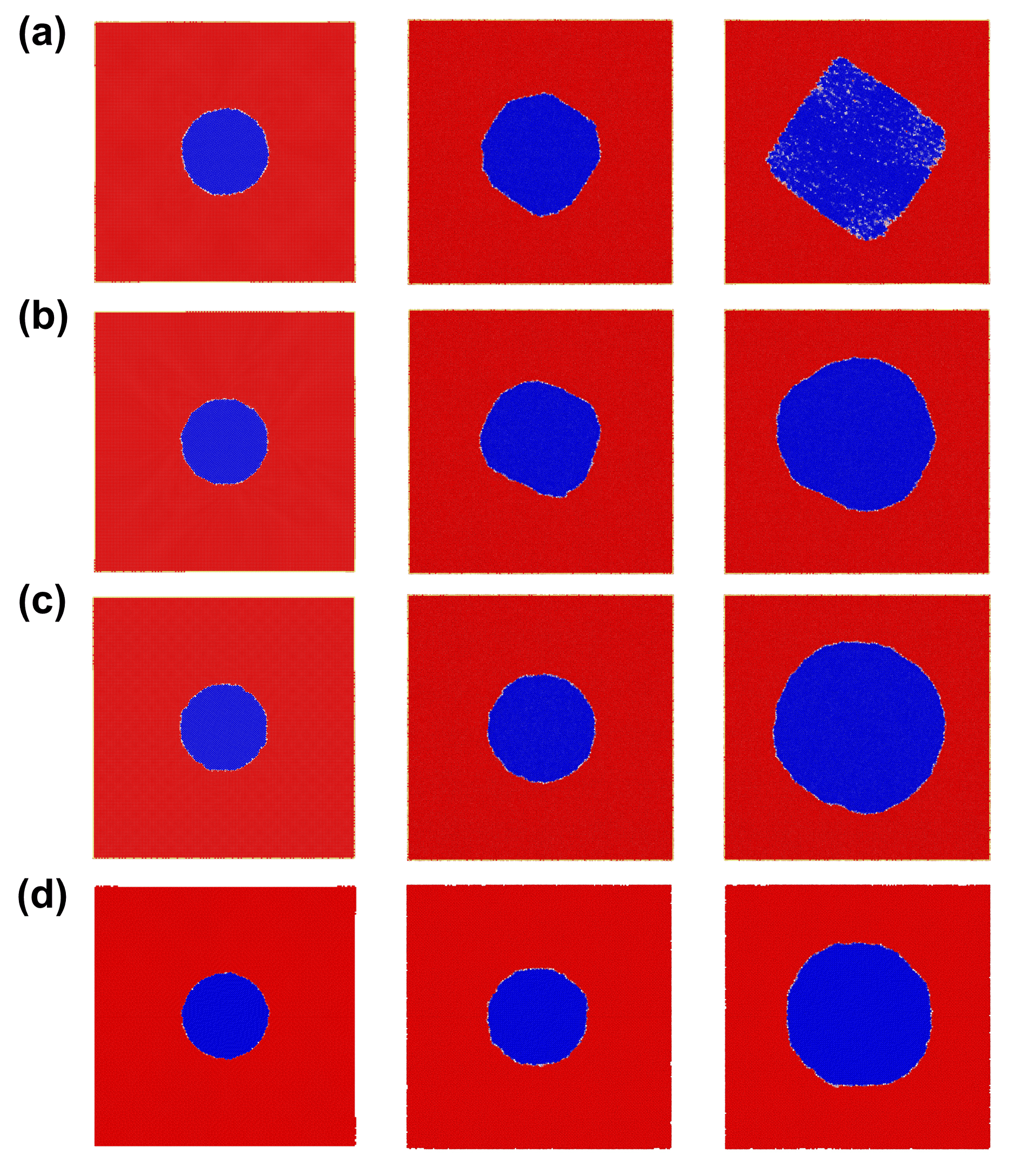}
    \caption{Snapshots of the expansion process of different unfixed boundaries at 1000K: (a) $\theta = 30 \degree$ boundary under SDF = 0.02 eV/$\ohm$ at simulations times of 0 ps, 90 ps, and 360 ps. (b) $\theta = 36.9 \degree$ boundary under SDF = 0.02 eV/$\ohm$ at simulaions times of 0 ps, 240 ps, and 420 ps. (c) $\theta = 40 \degree$ boundary under SDF = 0.026 eV/$\ohm$ at simulations times of 0 ps, 75 ps, and 270 ps. (d)$\theta = 45 \degree$ boundary under SDF = 0.026 eV/$\ohm$ at simulations times of 0 ps, 63 ps, and 237 ps. The atoms are colored by order parameter.}
    \label{fig:expand}
\end{figure}

Snapshots of the shape evolution process of the $\theta = 30 \degree$ unfixed boundary under SDF of 0.02 eV/$\ohm$, where $\ohm$ is the atomic volume, are shown in Figure \ref{fig:expand}(a). The atoms are colored according to order parameter. Red represents order parameter of 1 and blue represents -1. The snapshots show representative stages in the grain growth and faceting process. The inner grain grows from its initial cylindrical shape into an octagonal shape. Gradually, four of the facets in the octagonal shape shrink while the other four grow, transitioning the boundary into a square shape. The presence of atoms with ambiguous order parameters attests to the dynamic nature of the growth process. When the grain is faceted into a square shape, its growth is stagnated, and it no longer shrinks or expands. When different initial velocity seeds are applied for the simulations, the same facet can grow at slightly different rates, but the overall shape evolution process is the same.

%\begin{figure}[ht!]
%    \centering\leavevmode
%    \includegraphics[width=1\textwidth]{Figures/30_expand.png}
%    \caption{Snapshots of the expansion process of the $\theta = 30 \degree$ unfixed boundary at 1000 K under SDF = 0.02 eV/$\ohm$ at simulations times of (a) 0 ps (b) 90 ps (c) 360 ps. The atoms are colored by order parameter.}
%    \label{fig:30_expand}
%\end{figure}

The same type of grain growth and faceting behavior can be observed for the $15 \degree \leq \theta < 36.9 \degree$ boundaries: the shapes of the grain boundaries grow from cylinders to octagons, then squares. SDF cannot be applied to drive grain growth in low angle bicrystal systems ($\theta < 15 \degree$) because the order parameter does not sufficiently distinguish the two grains. For misorientations greater than $15 \degree$, the SDF method effectively identifies the two grains and drives grain growth and faceting.

The transition point between boundaries that form square shapes and boundaries that do not form distinct faceted shapes is around $\theta = 36.9 \degree$, which corresponds to a special coincidence site lattice (CSL) boundary. Snapshots of the grain growth process of the unfixed $\theta = 36.9 \degree$ boundary, also known as the $\Sigma5$ boundary, under a SDF of 0.02 eV/$\ohm$ are shown in Figure \ref{fig:expand}(b). For this boundary, distinct facets are formed during the initial stages of expansion, but the boundary eventually grows into an overall cylindrical shape, with transient facets that form and disappear quickly. The growth of the cylindrical grain is continuous under the applied driving force and stagnation does not occur.

%\begin{figure}[ht!]
%    \centering\leavevmode
%\includegraphics[width=1\textwidth]{Figures/36.87_expand.png}
%   \caption{Snapshots of the expansion process of the $\theta = 36.9 \degree$ unfixed boundary at 1000 K under SDF = 0.02 eV/$\ohm$ at simulaions times of (a) 0 ps (b) 240 ps (c) 420 ps.}
    %\label{fig:36.87_expand}
%\end{figure}

For the $\theta > 36.9 \degree$ boundaries, they no longer form distinct facets and continuously assume overall cylindrical shapes during expansion, as is shown in Figure \ref{fig:expand}(c) for the $\theta = 40 \degree$ unfixed boundary. In the expansion process, though the grain shape is not fully faceted, there are some dynamic facets that form and disappear quickly.

%\begin{figure}[ht!]
%    \centering\leavevmode
    %\includegraphics[width=1\textwidth]{Figures/40_expand.png}
    %\caption{Snapshots of the expansion process of the $\theta = 40 \degree$ unfixed boundary at 1000 K under SDF = 0.026 eV/$\ohm$ at simulations times of (a) 0 ps (b) 75 ps (c) 270 ps.}
    %\label{fig:40_expand}
%\end{figure}

The $\theta = 45 \degree$ boundary is a special boundary with different faceting behavior from the other investigated boundaries. While it does not form a fully faceted shape, several facets are formed multiple times during the simulation. The initial structure, along with two snapshots at which the inner grain has formed multiple facets while the overall shape is cylindrical, are shown in Figure \ref{fig:expand}(d). Eight facets with (100), ($\bar{1}$00), (010), (0$\bar{1}$0), (110), (1$\bar{1}$0), ($\bar{1}$10), and ($\bar{1}\bar{1}$0) orientations have been observed to form multiple times during the course of the simulation.

%\begin{figure}[ht!]
%    \centering\leavevmode
    %\includegraphics[width=1\textwidth]{Figures/45_expand.png}
    %\caption{Snapshots of the expansion process of the $\theta = 45 \degree$ unfixed boundary at 1000 K under SDF = 0.026 eV/$\ohm$ at simulations times of (a) 0 ps (b) 63 ps (c) 237 ps.}
    %\label{fig:45_expand}
%\end{figure}

\begin{figure}[ht!]
    \centering\leavevmode
    \includegraphics[width=1\textwidth]{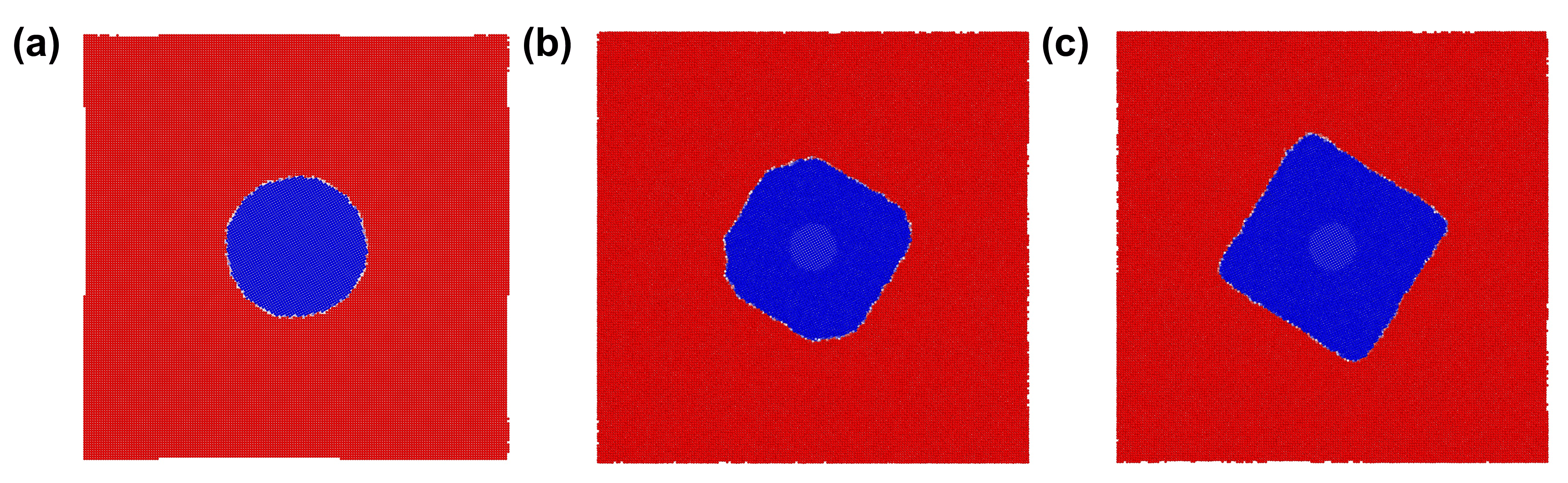}
    \caption{Snapshots of the expansion process of the $\theta = 30 \degree$ fixed boundary at 1000 K under SDF = 0.026 eV/$\ohm$ at simulations times of (a) 0 ps (b) 60 ps (c) 600 ps.}
    \label{fig:30_fixed_expand}
\end{figure}

When a small group of atoms in the center of the cylindrical grain is fixed to eliminate grain rotation, the shapes formed during grain growth are the same as for the unfixed boundaries. Snapshots of the expansion process of the $\theta = 30 \degree$ fixed boundary are shown in Figure \ref{fig:30_fixed_expand}. The octagonal and square shapes that are formed for the $\theta = 30 \degree$ unfixed boundary, are also formed for its fixed counterpart. Fixed boundaries with misorientations $15 \degree \leq \theta < 36.9 \degree$ also form octagonal and square shapes. The fact that the same faceted shapes are formed for the same initial misorientation, whether or not the cylindrical grain is free to rotate, implies that it is the initial misorientation that determines the faceted shape of the boundary. Less growth-related atomic disorder is observed in fixed grains, compared to unfixed grains, according to Figure \ref{fig:expand}(a) and Figure \ref{fig:30_fixed_expand}.

\begin{figure}[ht!]
    \centering\leavevmode
    \includegraphics[width=1\textwidth]{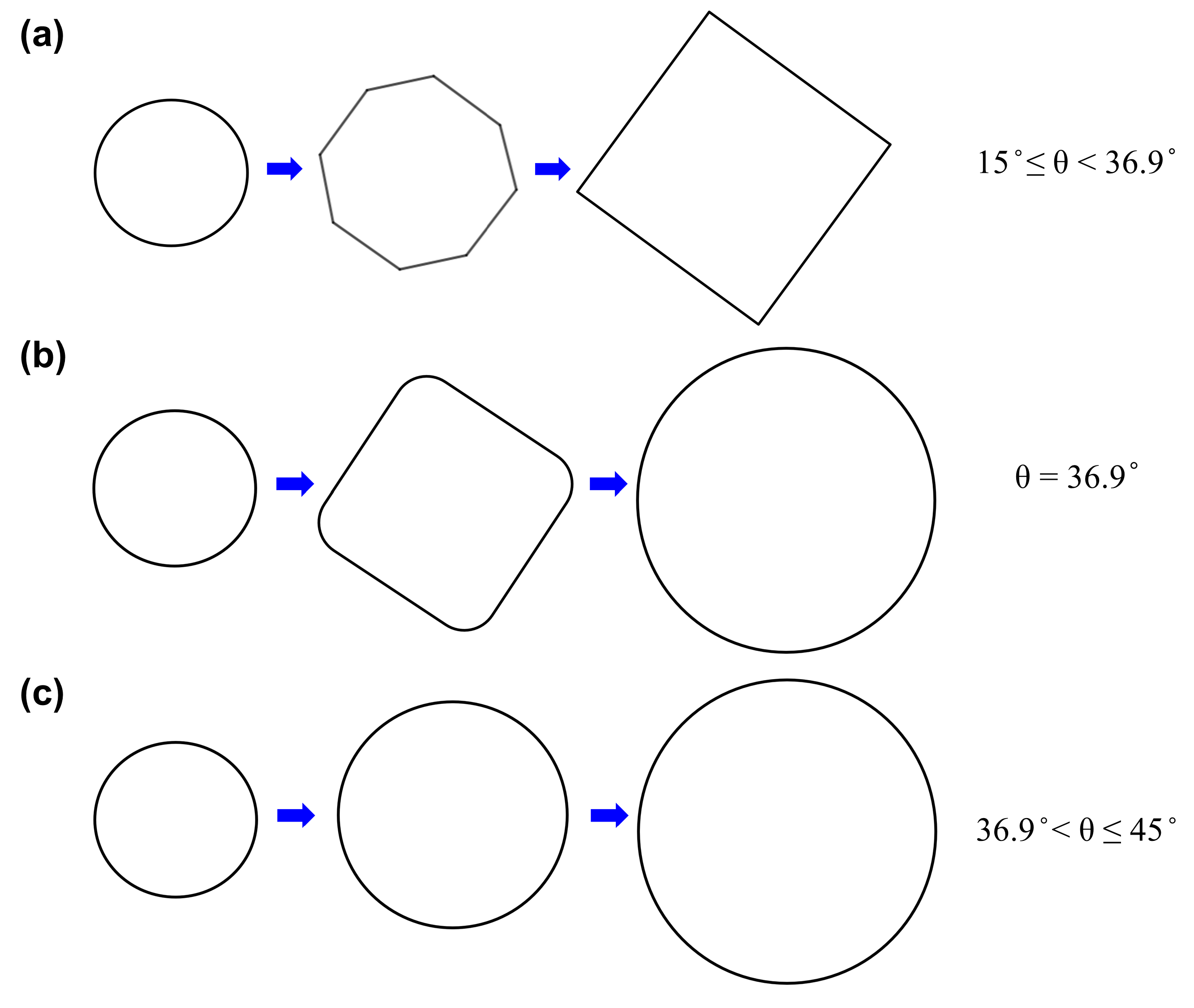}
    \caption{Shape evolution of cylindrical [001] tilt grain boundaries under applied synthetic driving force.}
    \label{fig:shape-evolution}
\end{figure}

Schematics of the shape evolution process during SDF driven expansion, which apply to both fixed and unfixed boundaries, are shown in Figure \ref{fig:shape-evolution}. For the same initial structure, the application of different initial velocity seeds may cause the same facet to grow at slightly different rates, but the overall shape evolution process is the same.

\newpage
\subsection{Grain rotation}

Grain rotation has been widely observed in the spontaneous shrinkage process of isolated cylindrical grains without external driving force in molecular dynamics simulations \cite{upmanyu_simultaneous_2006,trautt_grain_2012,molodov_grain_2015,french_molecular_2022,qiu_variability_2023}. Therefore, it is expected that grain rotation will also occur during expansion. The changes in misorientation during SDF-induced expansion are measured using quaternions. The method for measuring grain rotation, as well as the measured grain rotation rates in shrinking cylindrical grains, are described in detail in our previous publication \cite{qiu_variability_2023}. 

\begin{figure}[ht!]
    \centering\leavevmode
    \includegraphics[width=\textwidth]{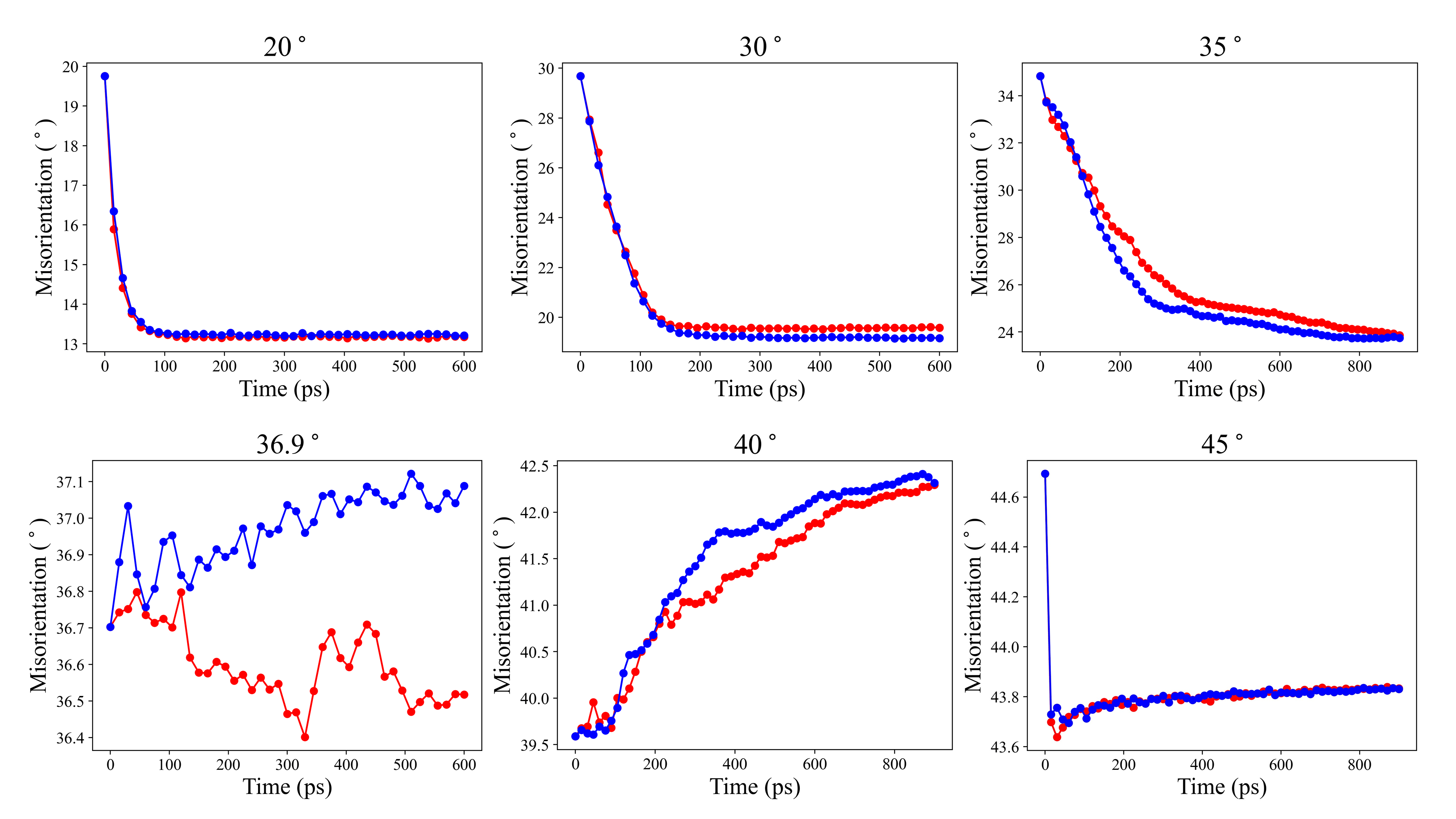}
    \caption{Misorientation vs. time plots for expanding cylindrical grains with initial misorientations 20-45 $\degree$. Two independent runs are are shown in blue and red.}
    \label{fig:rotation_expand}
\end{figure}

For each initial misorientation, two identical simulation runs differing only in initial velocity seeds are performed for grain rotation measurement under the same magnitude of SDF, to account for the possible variability in grain boundary motion that attributes to the different distributions of initial velocities. The misorientation vs. time plots for the $20 \degree \leq \theta \leq 45 \degree$ boundaries are shown in Figure \ref{fig:rotation_expand}. The applied SDF magnitude is 0.02-0.026 eV/$\ohm$ for all of the investigated boundaries, and each simulation runs for 600-900ps.

The $\theta = 36.9 \degree$ grain, which has been shown to exhibit minimal rotation during shrinkage, also exhibits very little rotation during expansion. The $36.9 \degree$ misorientation is near the critical angle for the reversal of the direction of rotation, for both shrinking and expanding boundaries, and it evinces both rotations. 

Compared to the shrinkage process, the expansion process reverses the grain rotation direction. Rotation to lower misorientations occurs in boundaries with initial misorientations less than $36.9 \degree$, while rotation to higher misorientations occurs in boundaries with initial misorientations greater than $36.9 \degree$, with the exception of the $\theta = 45 \degree$ boundary, in which the misorientation drops to around $43.6 \degree$ and only increases slightly over time. The initial velocity seeds impact the time evolution of rotation slightly, though not significantly.

\newpage
\subsection{Reversibility of faceting}

We have observed the expansion and faceting of cylindrical grain boundaries with certain initial misorientations under applied driving force. For boundaries with initial misorientations $15 \degree \leq \theta < 36.9 \degree$, stagnated growth can occur when the boundary is faceted into a square shape. At this stage, the inner grain ceases to expand and rotate. We refer to this stage as a quasi-equilibrium stage during which the grain has stopped evolving. 

\begin{figure}[ht!]
    \centering\leavevmode
    \includegraphics[width=1\textwidth]{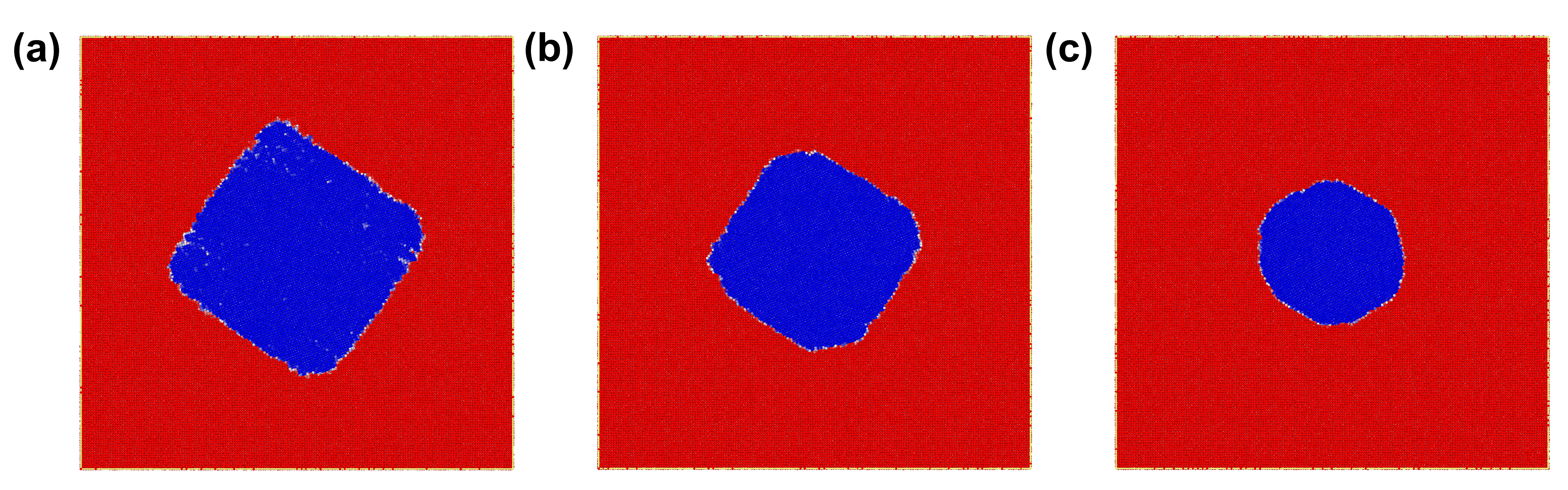}
    \caption{Shrinking grain for $\theta = 30 \degree$ unfixed boundary at 1000 K after the removal of SDF = 0.02 eV/$\ohm$ at different stages (a) removal of SDF (b) reappearance of facets associated with <100> dislocations (c) reformation of cylindrical grain boundary.}
    \label{fig:30_reverse}
\end{figure}

To test whether the square shape is stable without the SDF, the inner cylindrical grain is first expanded into a square shape under a SDF, and then the SDF is removed and the system is left to evolve on its own. As soon as the SDF is removed, the four facets in the square grain immediately begin to shrink, and eventually, four additional facets reappear, reforming the octagon shape. At some point, all well-defined facets disappear, and the cylindrical shape is reformed. The shape of the grain remains cylindrical until it vanishes. Figure \ref{fig:30_reverse} shows snapshots for the $\theta = 30 \degree$ unfixed boundary's evolution after the SDF is removed. In this process, the misorientation of the inner grain also starts to increase when the SDF is removed.

The shrinkage process of the reformed cylindrical grain is not the same as an initially cylindrical grain, as it has a much higher tendency to form transient facets during migration. The applied SDF has added energy to the system, so the system's free energy is increased. The faceting of a cylindrical grain boundary is reversible, though the path travelled by the shrinking boundary is not exactly the reverse of the path it travelled during expansion.

\newpage
\subsection{Facet structures and orientations}

\begin{figure}[ht!]
    \centering\leavevmode
    \includegraphics[width=\textwidth]{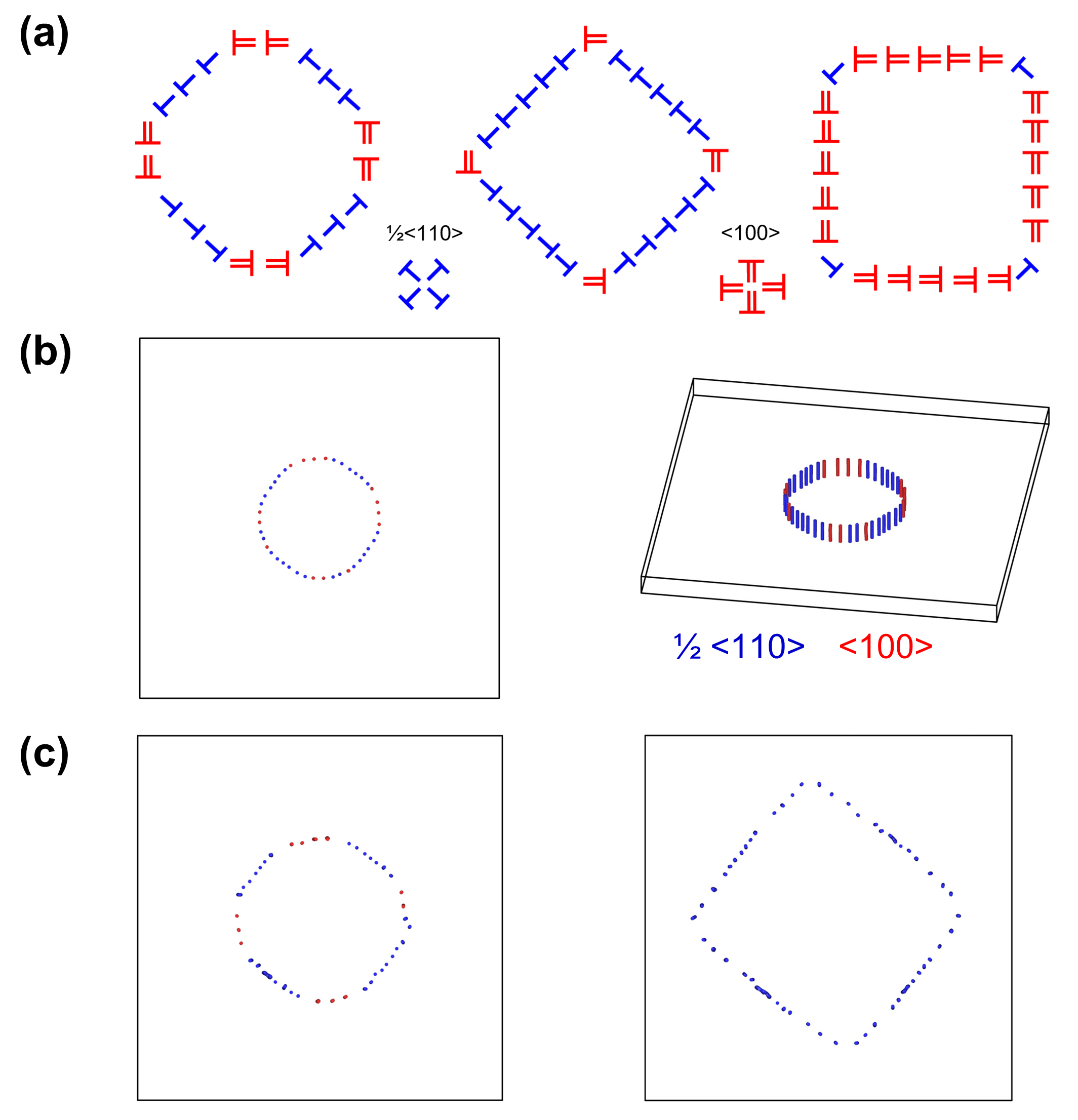}
    \caption{(a) Predicted faceted shapes of [001] cylindrical tilt grain boundaries and their dislocation compositions \cite{trautt_grain_2012} (b) Dislocations in the initially cylindrical $\theta = 20 \degree$ grain boundary, viewed from the x-y plane and side (c) Faceted octagonal and square shapes, viewed from the x-y plane.}
    \label{fig:dislocation-cylinder}
\end{figure}

Trautt and Mishin predicted the faceted shapes of cylindrical [001] tilt grain boundaries in Cu, which have FCC crystal structures similar to Ni \cite{trautt_grain_2012}. The predicted faceted shapes formed by two types of dislocations are shown in Figure \ref{fig:dislocation-cylinder}(a). The kite-shaped structural units in the grain boundary represent the cores of threading dislocations. The two types of dislocations, $\frac{1}{2}$<110> and <100>, are consistent with the results output by Ovito's dislocation analysis tool, which implements the Dislocation Extraction Algorithm (DXA) \cite{stukowski_automated_2012}. Figure \ref{fig:dislocation-cylinder}(b) shows the dislocations identified by Ovito's DXA modifier in the initially cylindrical $\theta = 20 \degree$ boundary. The facets that disappear when the square shape is formed are mainly composed of <100> type dislocations and the remaining facets are mainly composed of $\frac{1}{2}<110>$ type dislocations. The facets consisting of <100> type dislocations occur around $0 \degree$ boundary plane inclinations, while the facets consisting of $\frac{1}{2} <110>$ type dislocations occur around $45 \degree$ inclinations. Note that the dislocations identified here correspond to the kite-shaped structural units in the grain boundary, not the dislocation component of the disconnections.

No dislocations have been identified for $\theta >= 30 \degree$ boundaries using DXA in OVITO, possibly due to dislocation cores merging at higher misorientations. However, since the faceted shapes of the $\theta = 20 \degree$ and $\theta >= 30 \degree$ boundaries are similar, we assume that the dislocations structures in the $\theta >= 30 \degree$ boundaries are similar to that of the $\theta = 20 \degree$ boundary.

Of the three predicted shapes, octagon formed by both types of dislocations, square formed by $\frac{1}{2}$<110> type dislocations, and square formed by <100> type dislocations, only the first two have been observed from our simulations of growing cylindrical grains driven by SDF. For the boundaries with initial misorientations of $15 \degree \leq \theta < 36.9 \degree$, facets consisting of <100> dislocations shrink while facets consisting of $\frac{1}{2}$<110> dislocations grow.

For each facet, there is very little change in the orientation during the migration process. For the $\theta = 20 \degree$ boundary, the orientations of the facets corresponding to the $\frac{1}{2}$<110> dislocations are (4$\bar{3}$0), ($\bar{3}\bar{4}$0), (340), and ($\bar{4}$30). The approximate orientations of the facets corresponding to the <100> type dislocations are (1$\bar{4}$0), ($\bar{4}\bar{1}$0), (410), and ($\bar{1}$40). The facet orientations are consistent with the four fold symmetry of the boundary. In the $\theta = 30 \degree$ boundary, the facets corresponding to the $\frac{1}{2}$<110> dislocations are approximately (3$\bar{2}$0), ($\bar{2}\bar{3}$0), (230), and ($\bar{3}$20); the approximate orientations of the facets corresponding to the <100> type dislocations are (1$\bar{5}$0), ($\bar{5}\bar{1}$0), (510), and ($\bar{1}$50). The orientations of the facets formed in the two boundaries are similar enough that it can be difficult to distinguish between them, especially during the migration process in which there are thermal fluctuations. The other boundaries with initial misorientations $15 \degree \leq \theta < 36.9 \degree$ also have similar facet orientations. For the unfixed boundaries, the facets in the square shapes are close to symmetric tilt boundaries. 

The facet orientations at various temperatures between 700 K and 1400 K, for both unfixed and fixed boundaries, are also investigated. For the same initial misorientation, the facet orientations are found to be the same, regardless of temperature and whether or not rotation is prohibited. Varying the radius of the cylindrical grain from 11$a_0$ to 30$a_0$, the same facet orientations can also be observed, as long as the initial misorientation is the same.

\newpage
\subsection{Mechanisms of facet growth}

\begin{figure}[ht!]
    \centering\leavevmode
    \includegraphics[width=1\textwidth]{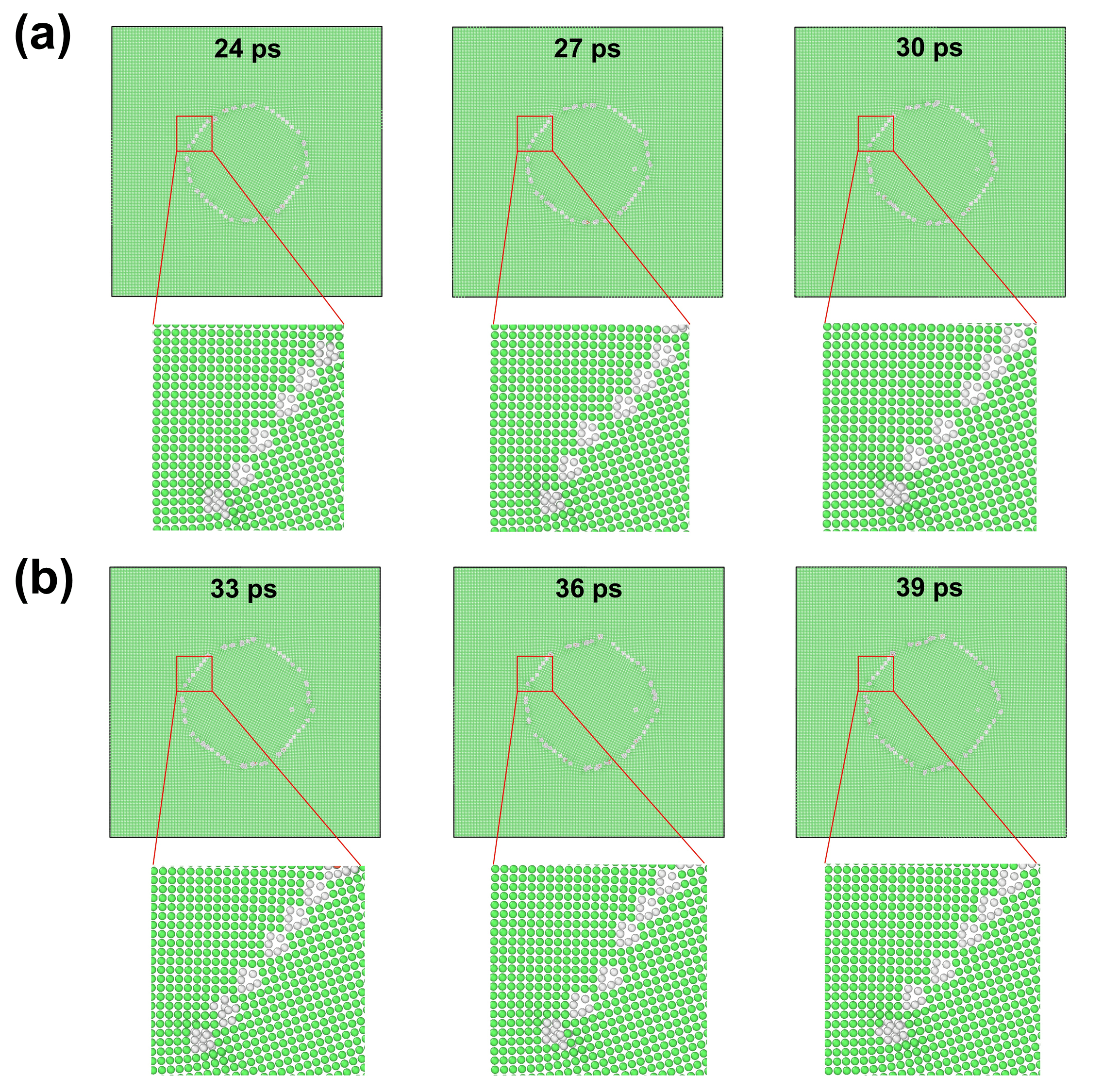}
    \caption{Migration of a facet by the nucleation and motion of two disconnections in the $\theta = 20 \degree$ unfixed boundary under SDF of 0.02 eV/$\ohm$ at (a) 24-30ps (b) 33-39ps.}
\label{fig:disconnection-step}
\end{figure}

The facets in the grain boundary structure migrate by the nucleation and motion of disconnections. Figure \ref{fig:disconnection-step} shows the migration process of a facet in the grain boundary structure mediated by the nucleation and motion of two disconnections at different times in the $\theta = 20 \degree$ unfixed boundary, which is composed of discrete kite-shaped structural units. During facet migration, a portion of the kite-shaped units in the facet will move forward first, forming a step in the facet structure, and then the rest of facet will follow. This process is the nucleation and migration of a disconnection. Figure \ref{fig:disconnection-step}(a) shows facet migration at simulations times of 24-30ps through the nucleation and migration of a disconnection, and Figure \ref{fig:disconnection-step}(b) shows a different disconnection at 33-39ps. Each snapshot shown in Figure \ref{fig:disconnection-step} undergoes energy minimization through MS simulations to remove thermal noise, and the atoms are colored using PTM.

Disconnections can nucleate at various positions in the grain boundary at different times. Multiple disconnections can nucleate at the same time within the same facet. The nucleation rate of disconnections is much faster in the facets consisting of <100> type dislocations, so the migration rate is also faster compared to the facets consisting of $\frac{1}{2}$<110> type dislocations. The nucleation of disconnections is also affected by initial velocity seeds. For the same initial boundary structure under the same magnitude of SDF, in different simulation runs with different initial velocity seeds, the disconnection nucleation processes can be different, causing the same facets to migrate at slightly different rates. Therefore, it is difficult to determine all the possible disconnections associated with the migration of each facet through MD simulations. It is also difficult to characterize the geometries of the disconnections, with the complications from grain rotations, lattice distortions, and the shrinkage and growth of facets. 

For the grain boundaries that do not form well-defined faceted shapes ($\theta >= 36.9$), boundary migrations are also mediated by disconnections. In those boundaries, the kite-shaped structural units significantly overlap and are not easily distinguishable from each other. During boundary migration, steps are constantly forming at various sites in the grain boundary, and the formation sites of the steps are very close to each other, so that it is difficult to distinguish between them, causing the boundary to expand in an almost uniform manner. 

\newpage
\subsection{Wulff shape vs. kinetic shape}

\begin{figure}[ht!]
    \centering\leavevmode
    \includegraphics[width=1\textwidth]{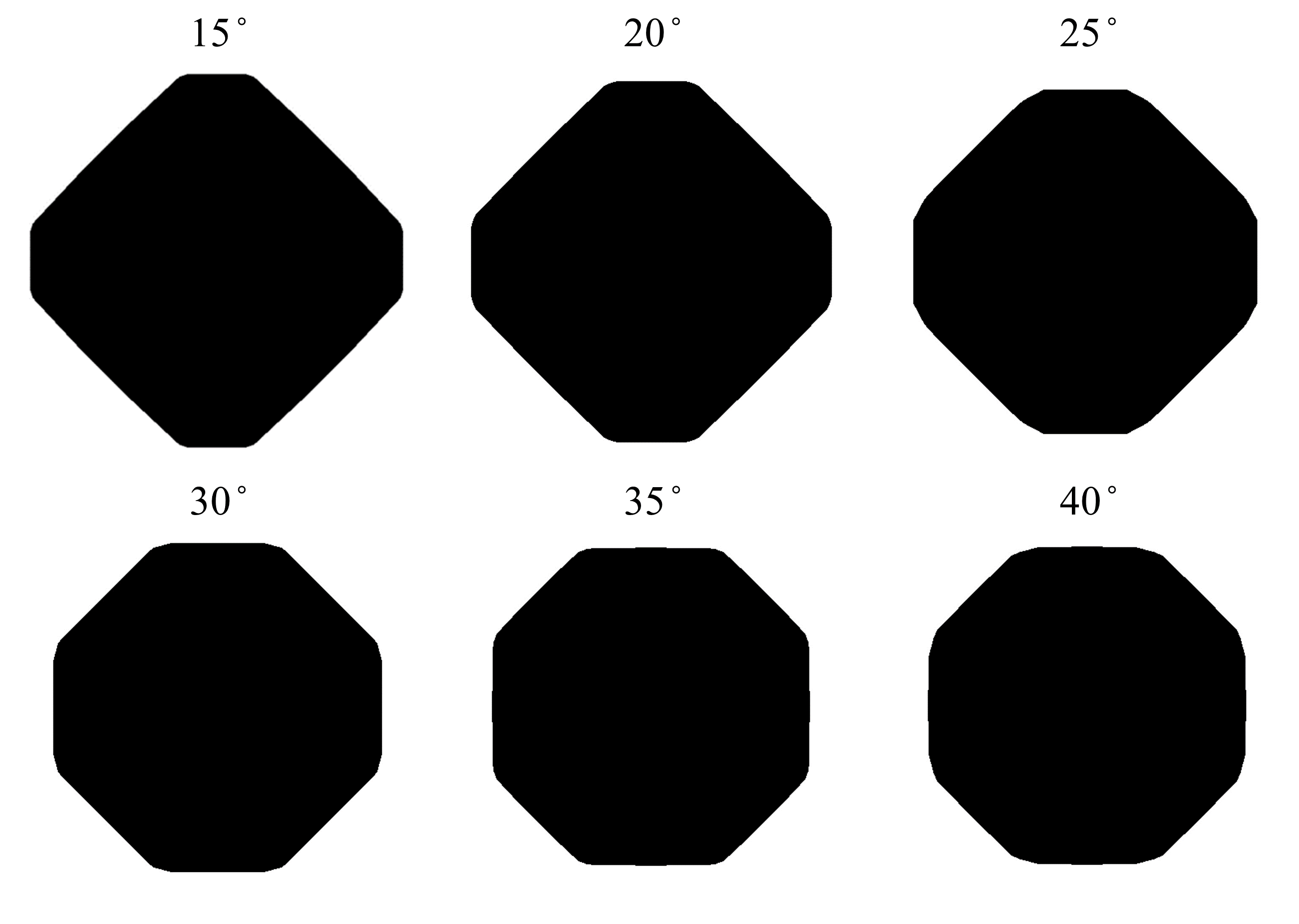}
    \caption{Wulff shapes of non-CSL $\theta = 15-40 \degree$ boundaries}
    \label{fig:Wulff-general}
\end{figure}

\begin{figure}[ht!]
    \centering\leavevmode
    \includegraphics[width=1\textwidth]{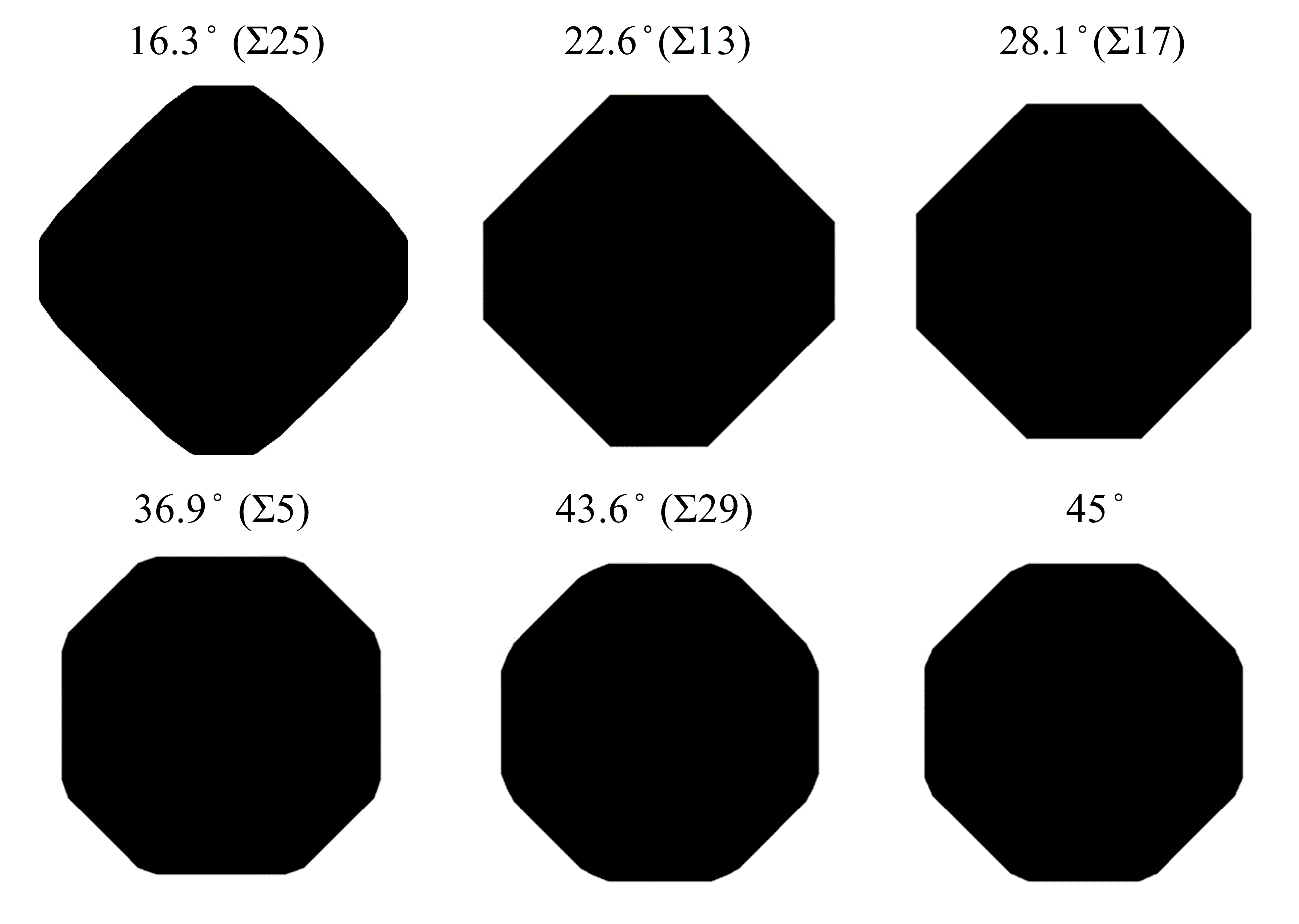}
    \caption{Wulff shapes of special grain boundaries: $\theta = 16.3 \degree, 22.6 \degree, 28.1 \degree, 36.9 \degree, 43.6 \degree, 45 \degree$}
    \label{fig:Wulff-special}
\end{figure}

The computed Wulff shapes of general grain boundaries that do not belong to any CSL are shown in Figure \ref{fig:Wulff-general}, and the Wulff shapes of special CSL grain boundaries are shown in Figures \ref{fig:Wulff-special}. For all of the investigated boundaries, facets exist at the $0 \degree$ and $45 \degree$ inclinations. The computed Wulff shapes are not the same as the kinetic shapes observed from the MD simulations. While the Wulff shapes are all octagonal, the kinetic shapes evolve from octagons to squares, or remain cylindrical. 

The Wulff shapes are purely derived from grain boundary energy anisotropy, while the kinetic shapes from MD simulations are mainly determined by grain boundary mobility anisotropy. This does not mean, however, that the kinetic shapes are not affected by grain boundary energy. For the $15 \degree \leq \theta < 36.9 \degree$ boundaries that form well-defined faceted shapes, the facets in the octagonal shapes are close to $0 \degree$ and $45 \degree$ boundary plane inclinations, which also correspond to the facets in the Wulff shapes.  

\clearpage
\subsection{Numerical simulations with continuum disconnection models}

In previous studies, we demonstrated that the flow of disconnections on the grain boundary during migration can lead to grain boundary faceting \cite{qiu_interface_2023} and grain rotation \cite{qiu_disconnection_2024}.
Due to the influence of internal stress caused by the disconnection motion, the kinetic shapes of the embedded grains during shrinkage/growth could certainly differ from the prediction based on Wulff construction (i.e., pure capillarity effect) found in our MD simulations.

\begin{figure}[ht!]
    \centering\leavevmode
    \includegraphics[width=0.8\textwidth]{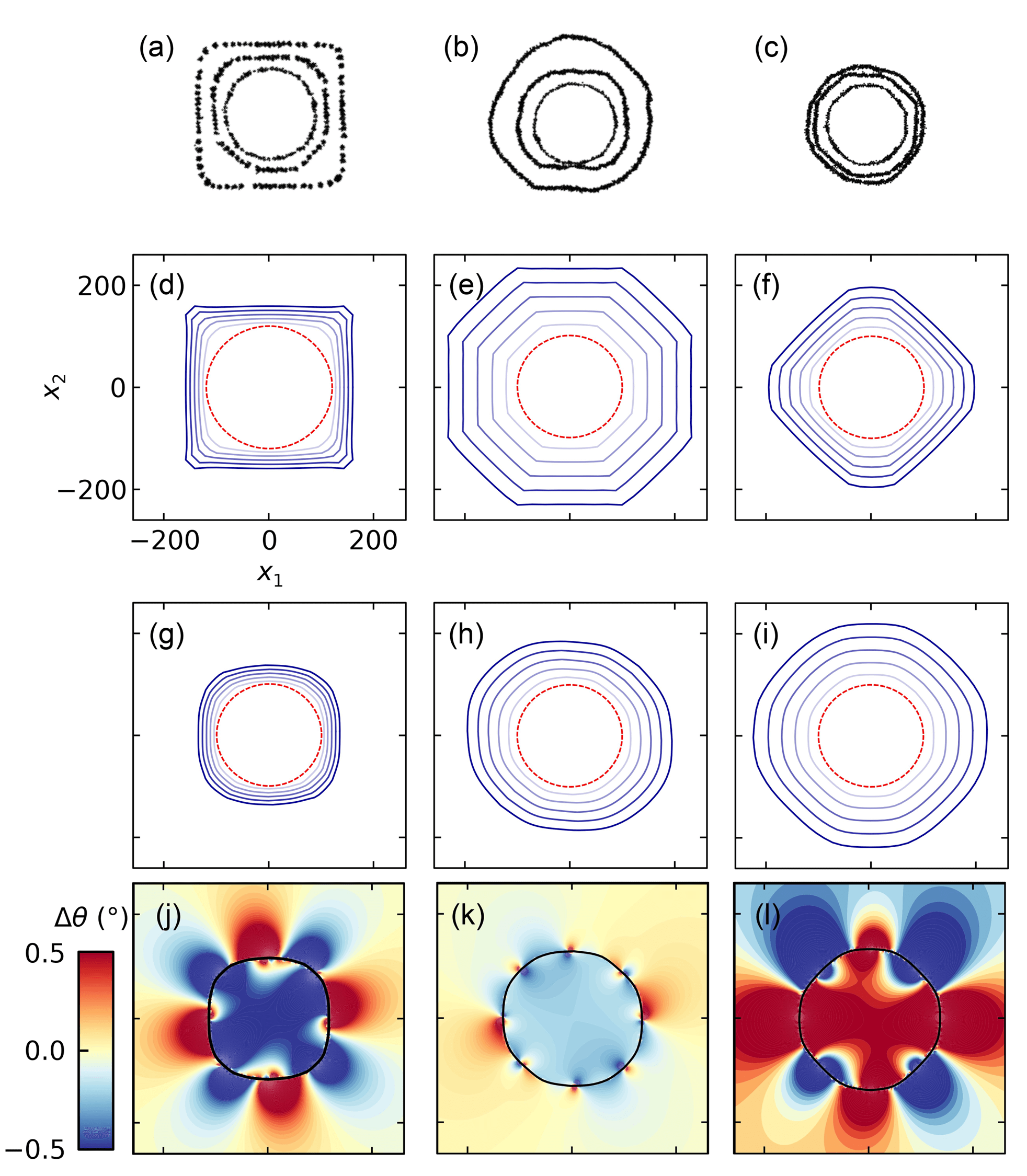}
    \caption{Shape evolution of $\theta = 28.1 \degree$ ($\Sigma17$ in (a), (d), (g), (j)), $\theta = 36.9 \degree$ ($\Sigma5$ in (b), (e), (h), (k)), and $\theta = 43.6 \degree$ ($\Sigma29$ in (c), (f), (i), (l)) boundaries. 
    (a)-(c) MD simulations results at representative stages of grain growth.  
    (d)-(f) Single-disconnection-mode model results with time interval $\Delta t = 1000$. 
    The red dash circle represents the initial grain boundary structure. 
    (g)-(i) Two-disconnection-mode model results with time interval $\Delta t = 1000$. 
    (j)-(l) Rotation fields $\Delta \theta$ at time $t = 2500$. 
    }
    \label{fig:shape-continuum}
\end{figure}

Figure~\ref{fig:shape-continuum} shows the comparisons about the kinetic shapes and grain rotation directions of the embedded grains with three typical misorientation angles discussed above (i.e., $\Sigma 17 (\theta = 28.1\degree), ~ \Sigma 5 (\theta =36.9\degree)$ and $\Sigma 29 (\theta =43.6\degree$) between the results obtained by MD simulations and continuum disconnection models.
For each of the three boundaries, boundaries in MD simulations are extracted and overlaid at representative grain growth stages, see Figures~\ref{fig:shape-continuum}(a)-(c). 
Note that, the grain boundaries are rotated by $\theta/2$ to align one facet horizontally and other three facets with inclination angles of $45\degree$, $90\degree$ and $135\degree$, in the same manner shown in the Wulff plots (e.g., see Figure \ref{fig:Wulff-construction}).

Figures~\ref{fig:shape-continuum}(d)-(f) provide the evolution of the initially-circular grain (the red dash lines) during grain growth driven by capillarity force, internal stress caused by disconnection motion and applied synthetic driving force.
Here, only \emph{one} disconnection mode with lowest activation energy is considered, see Table 2 in the Supplementary Information.
MD simulations and the numerical simulations based on the continuum model give exactly the same crystal orientations of the facets.
Especially, the kinetic shapes of the $\Sigma 17$ grain boundary obtained by both approaches match perfectly.
For $\Sigma 5$ and $\Sigma 29$ grain boundaries, kinetic shapes obtained by the continuum models are much more faceted than those in MD simulations.

Continuum model regarding multiple disconnection modes is more realistic for comparisons with MD simulations at relatively high temperatures ($T = 1000~\text{K} \approx 0.6 T_\text{melt}$).
In Figures~\ref{fig:shape-continuum}(g)-(i), two disconnection modes on each reference interface with low activation energies are considered, see Table 2 in the Supplementary Information. 
The initial ratio between two disconnection modes are set as -1 such that no internal long-range stress is generated at first (see more explanations in literature~\cite{qiu_disconnection_2024}).
Grain boundaries appear to be less faceted compared with the single-disconnection-mode results in Figs.~\ref{fig:shape-continuum}(d)-(f).
Also, we demonstrate that the larger the grain sizes are, the less faceted the grain boundaries will be, which is consistent with the MD results.
Both grain boundary energy anisotropy (i.e., capillarity force) and internal stress enhance grain boundary faceting.
As the grain grows, the capillarity force proportional to the grain boundary curvature and the internal stress inverse to the grain size will decrease.
Thus, the effect of the applied isotropic synthetic driving force will be more and more prominent to determine the grain boundary shape and lead to \emph{defaceting}.

Another difference between the single and multiple disconnection modes is that grain rotation and grain boundary sliding can only happen considering multiple disconnection modes.
In our previous study, grain rotation directions during \emph{shrinkage} of the $\Sigma 25, ~ \Sigma 17, ~ \Sigma 5$ and $\Sigma 29$ boundaries predicted by our continuum model are consistent with MD simulation results \cite{qiu_disconnection_2024}.
Here, we investigate the grain rotation directions of those boundaries during \emph{growth} under applied synthetic driving forces, see the rotation fields in Figures~\ref{fig:shape-continuum}(j)-(l).
Here, rotation field is extracted as the skew part of the displacement gradient generated by the disconnection Burgers vector densities along the grain boundary.
The embedded grains surrounded by $\Sigma 17$ and $\Sigma 5$ grain boundaries rotate to decrease misorientation, whilst the grain surrounded by $\Sigma 29$ grain boundary rotate to increase misorientation during grain growth (opposite to the rotation direction during grain shrinkage in \cite{qiu_disconnection_2024}).
We demonstrate that the grain rotation direction during grain growth is also consistent with current MD simulations shown in Figure \ref{fig:rotation_expand}.
We note that, in order to match the grain rotation rate, only one parameter in our continuum model needs to be fitted.
However, grain rotation direction can be fully determined without this fitting scheme (which is determined by bicrystallographic and thermodynamic materials parameters).

We find the consistency of the kinetic shapes and grain rotation directions between results obtained by current MD simulations and continuum disconnection models. 
Hence, disconnection flow/motion on the grain boundary during migration can be used to understand and predict the kinetic shapes of grain boundaries and grain rotation directions.

\newpage
\section{Conclusions}
The kinetic growth shapes of isolated cylindrical [001] tilt grain boundaries are investigated in MD simulations by applying a synthetic driving force to enable expansion. During this process, distinct faceted shapes may or may not be observed, depending on the initial misorientation. 

Cylindrical grain boundaries with initial misorientations $15 \degree \leq \theta < 36.9 \degree$ are capable of forming two types of facets consisting of <100> type and $\frac{1}{2}$<110> type dislocations. Initially, an octagonal shape consisting of both types of facets are formed. Then the facets composed of <100> type dislocations shrink and facets composed of $\frac{1}{2}$<110> type dislocations expand, eventually transforming into a square shape. The orientations of the facets remain overall stable during migration, and are only found to be affected by initial misorientation. Other factors such as driving force magnitude, temperature, grain radius, and grain rotation have little effect on the orientations of the facets. When the boundaries are unfixed, rotation of the inner cylindrical grains to lower misorientations can be observed.

Cylindrical grain boundaries with initial misorientations $36.9 \degree < \theta \leq 45 \degree$ do not form well-defined faceted shapes, only transient facets. The grain shapes are cylindrical throughout the expansion process, and rotations to higher misorientations can be observed. The transition point between boundaries that form well-defined faceted shapes and boundaries that remain cylindrical is around $\theta = 36.9 \degree$. The $\theta = 36.9 \degree$ boundary, which is a minimally rotating boundary that can form some facets, is also the transition point in the reversal of rotation direction.

The motions of facets are mediated by the nucleation and migration of disconnections. In the faceted octagonal shapes, the nucleation and migration of disconnections are faster in the shrinking facets than in the expanding facets, causing the boundary to transform into a square shape. The nucleation and motion of disconnections are stochastic, affected by different initial velocity seeds, which may result in slightly different migration rates for the same facet. 

The kinetic shapes of the cylindrical grain boundaries observed in MD simulations are not the same as the computed Wulff shapes. The kinetic shapes are bound by slowest moving grain boundary planes while the Wulff shapes are bound by lowest energy grain boundary planes. 

Comparisons of MD simulation results with numerical simulations using continuum disconnection models show that the migration of grain boundaries, as well as the associated grain rotations, are mediated by the flow of multi-mode disconnections along the grain boundaries. The rotation directions of the embedded grains surrounded by the $\theta = 28.1 \degree$ ($\Sigma17$), $\theta = 36.9 \degree$ ($\Sigma5$), and $\theta = 43.6 \degree$ ($\Sigma29$) grain boundaries are accurately captured by the two-disconnection-mode model. In some cases, the kinetic shapes of the boundaries from MD simulations match the kinetic shapes from the numerical simulations very well. Overall, the faceting tendency predicted by MD simulations and the continuum model are consistent.

The growth and faceting of isolated cylindrical grains under synthetic driving force is analogous to abnormal grain growth in polycrystals. During abnormal grain growth, some large grains expand quickly against curvature in a matrix of many much smaller grains that grow at much slower rates. The abnormal grains move due to stored energy, similar to the applied SDF, and are often highly faceted \cite{lee_grain_2000,zhao_abnormal_2015}. The investigations of the grain growth and faceting mechanisms of isolated cylindrical grains can also provide guidelines for the growth and faceting of abnormal grains in polycrystals.

\section*{Acknowledgments}

AQ and EAH were supported in part by National Science Foundation grants DMR-1710186 and DMR-2118945.  DJS was supported by the Research Grants Council, Hong Kong SAR through the General Research Fund (17210723). We would like to thank Y. Mishin and T. Frolov for helpful discussions.

\newpage
This is the Supplementary Information of paper "Kinetic and Equilibrium Shapes of Cylindrical Grain Boundaries".
%\begin{frontmatter}
\title{Supplementary Information}

\section{Quenching structures from high temperature}

It is difficult to determine the facet orientations using raw simulation data, as there are a lot of thermal fluctuations at our chosen simulation temperatures. The initial grain boundary structures at 0 K can be identified using polyhedral template matching (PTM). The initial structure of the $\theta = 20 \degree$ boundary, for example, consists of multiple discrete empty kite shapes. At high temperatures, when attempting to use PTM to identify grain boundary structures, many regions in the perfect crystals are identified as defected regions. While the grain boundary region can be roughly identified, not much structural information can be obtained.

\begin{figure}[ht!]
    \centering\leavevmode
    \includegraphics[width=1\textwidth]{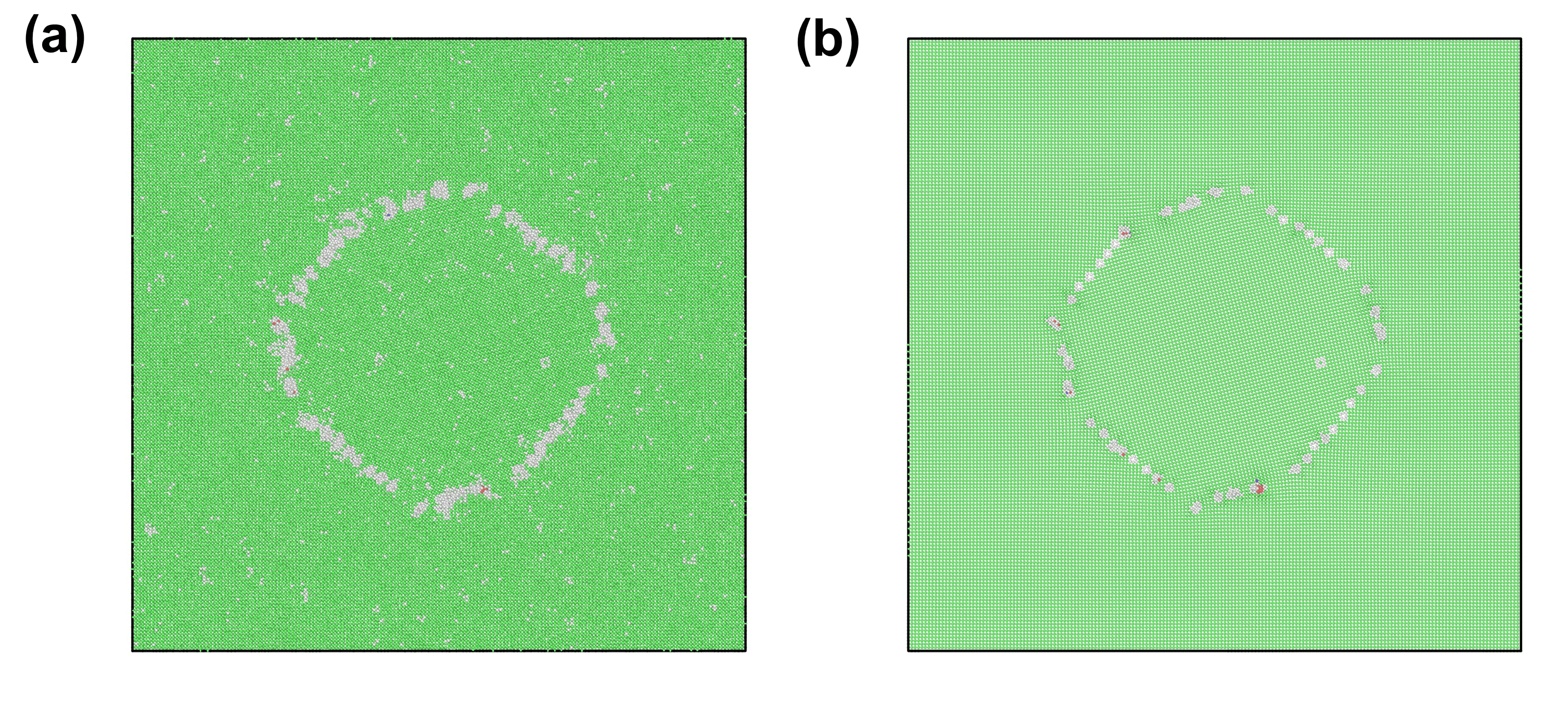}
    \caption{Snapshot of $\theta = 20 \degree$ free boundary with octagonal shape (a) before (b) after energy minimization to remove thermal fluctuations.}
    \label{fig:disorder-order}
\end{figure}

In order to remove the thermal fluctuations, the simulation snapshots containing atom information are optimized by molecular statics using the Polak-Ribiere version of the conjugate descent algorithm \cite{hestenes_methods_1952} at 0 K. By performing the energy minimization step, the thermal noise can be effectively removed. Supplementary Figure \ref{fig:disorder-order} shows a simulation snapshot of the $\theta = 20 \degree$ boundary under an applied driving force of 0.02 eV/$\ohm$ after 45ps of simulation time, before and after energy minimization. In this snapshot, the shape of the inner grain is an octagon. After energy minimization, ordered kite shapes can be identified within the grain boundary structure. There are multiple empty and filled kite shapes within the grain boundary, and the facets corresponding to $<100>$ type dislocations consist of filled kites while the facets corresponding to $\frac{1}{2}<110>$ type dislocations consist mostly of empty kites.

Snapshots of the $\theta = 30 \degree$ boundary at different faceting stages after energy minimization are shown in Supplementary Figure \ref{fig:30-poly-mini}. The initial structure of the $\theta = 30 \degree$ boundary consists of interconnected empty kite shapes. Though the kites are interconnected, each is distinguishable from its neighbors. As the boundary expands under SDF, some of the interconnected kite shapes begin to separate from each other. When the boundary is faceted, the kite shapes in the grain boundary structure are also more connected compared to the $\theta = 20 \degree$ boundary.

\begin{figure}[ht!]
    \centering\leavevmode
    \includegraphics[width=1\textwidth]{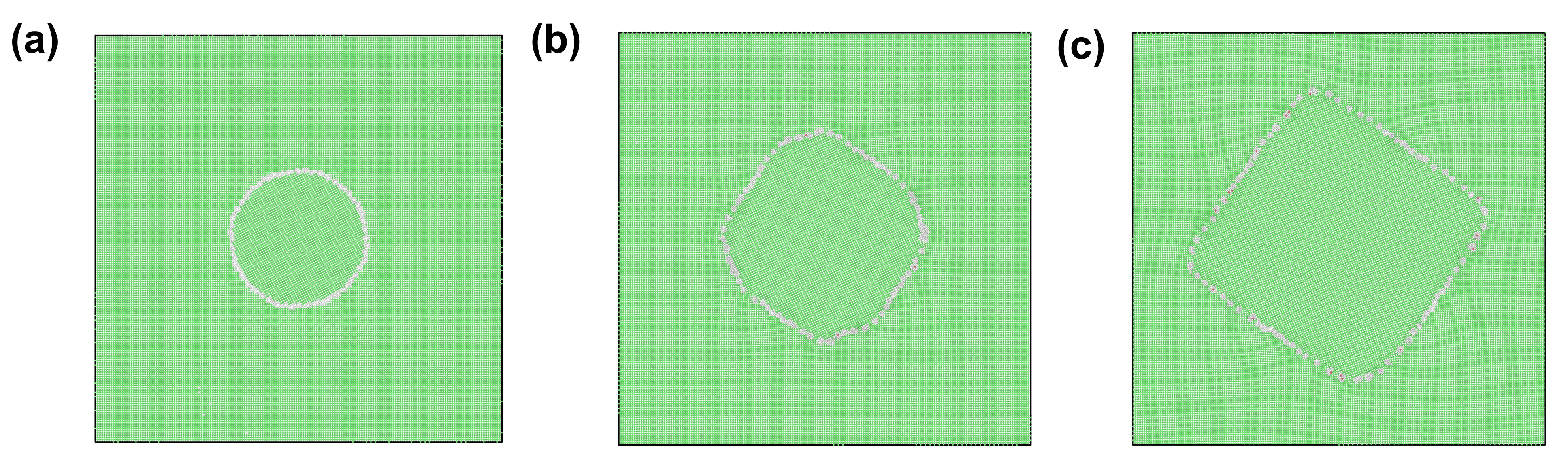}
    \caption{Snapshots of $\theta = 30 \degree$ unfixed boundary at different stages of faceting (a) cylindrical shape before simulation run (b) octagon (c) square}
    \label{fig:30-poly-mini}
\end{figure}

\begin{figure}[ht!]
    \centering\leavevmode
    \includegraphics[width=1\textwidth]{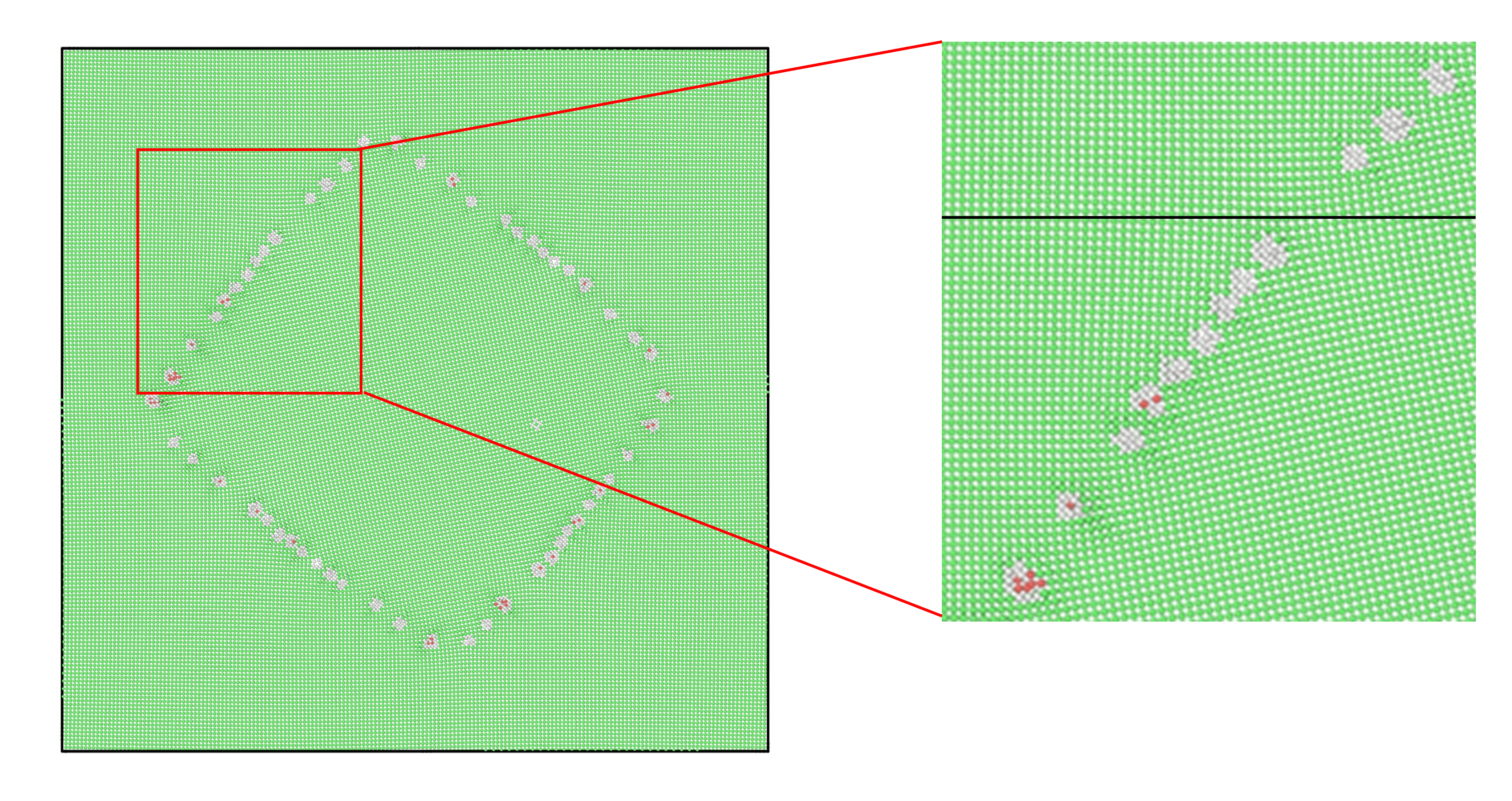}
    \caption{Magnified facets region in the square shape formed by the $\theta = 20 \degree$ unfixed boundary.}
    \label{fig:symm}
\end{figure}

For the unfixed boundaries, the facets in the square shape have the tendency to become symmetric tilt boundaries, accommodated by the rotation of the inner grain and slight distortions in the lattices of the outer matrix grains. Supplementary Figure \ref{fig:symm} shows the magnified version of a facet region in a $\theta = 20 \degree$ boundary when a square shape is formed. There are some lattice distortions around the facet region, as indicated by the black horizontal line. For the fixed boundaries, the fixed core region applies a torque to prevent grain rotation, but there are still some rotations in the inner grain as the facets strive to become symmetric tilt boundaries.

\begin{figure}[ht!]
    \centering\leavevmode
    \includegraphics[width=1\textwidth]{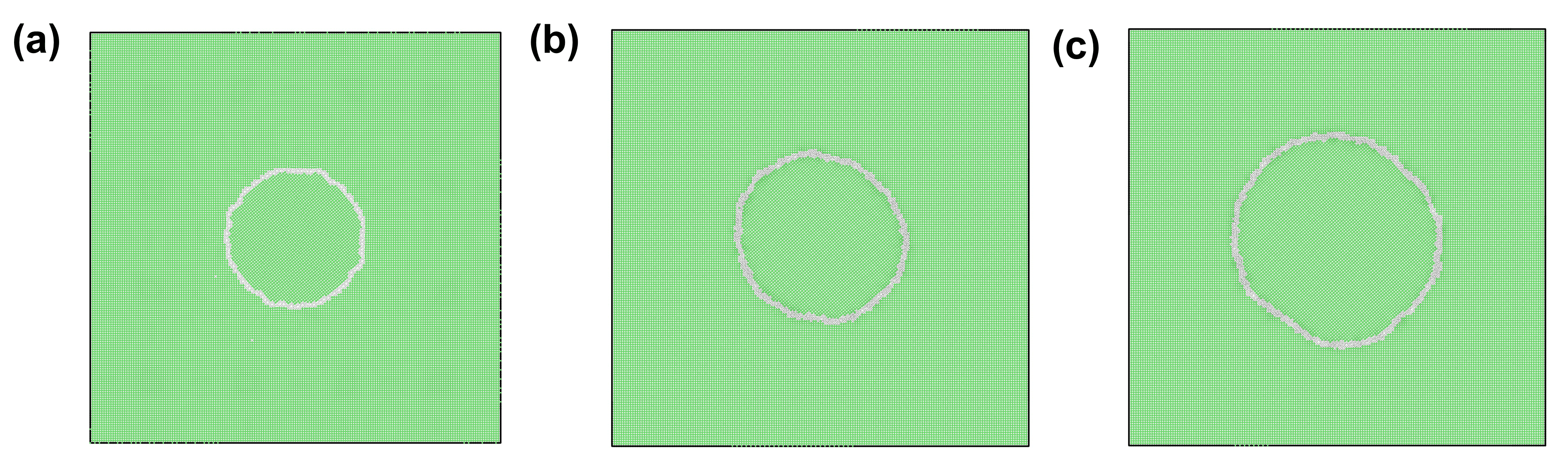}
    \caption{Snapshots of $\theta = 40 \degree$ unfixed boundary at simulation times under a SDF of 0.02 eV/$\ohm$ (a) 0ps (b) 150ps (c) 300ps.}
    \label{fig:40-poly-mini}
\end{figure}

In the initial $\theta = 40 \degree$ boundary, most of the kite shaped structures are indistinguishable from each other. As the boundary expands, most of the kites are still indistinguishable from their neighbors. The growth of the boundary is isotropic and no distinct facets are formed. Supplementary Figure \ref{fig:40-poly-mini} shows snapshots of the migration process of the $\theta = 40 \degree$ boundary after energy minimization.

\newpage
\section{Details of the multiple disconnection mode model}

\subsection{Geometry, disconnection description and relaxation of constraint}

Building on the continuum description introduced in Refs.~\cite{han2022disconnection,sal2022disconnection}, the grain boundaries (GBs) can be described by curves in a $\mathbf{e}_1 - \mathbf{e}_2$ Cartesian coordinate system:
\begin{equation}
\mathbf{x}(s) = \left(\begin{array}{ccc}
	x_1(s) \\
	x_2(s)
	\end{array}\right) ,
\end{equation}
where $s$ is a parameter; e.g. if $s$ is time, $\mathbf{x}(s)$ represents a trajectory.
If $s$ is inclination angle $\phi$, $\mathbf{x}(s)$ refers to the coordinate of the point on the GB with inclination angle $\phi$.
The inclination angle of a GB segment can be calculated by the unit tangent vector $\mathbf{l}(s)$ at this point as
\begin{equation}
	\mathbf{l}
	=
	\frac{\mathbf{l}}{\vert{\mathbf{l}}\vert}
	=
	\frac{1}{\vert{\mathbf{l}}\vert}\frac{{\rm d}\mathbf{x}}{{\rm d}s}
	=
	\frac{{\rm d}\mathbf{x}}{{\rm d}L}
	=
	\left(\begin{array}{ccc}
	l_1 \\
	l_2
	\end{array}\right) 	.
\end{equation}

\begin{figure}[tb]
\centering
\includegraphics[width=0.6\linewidth]{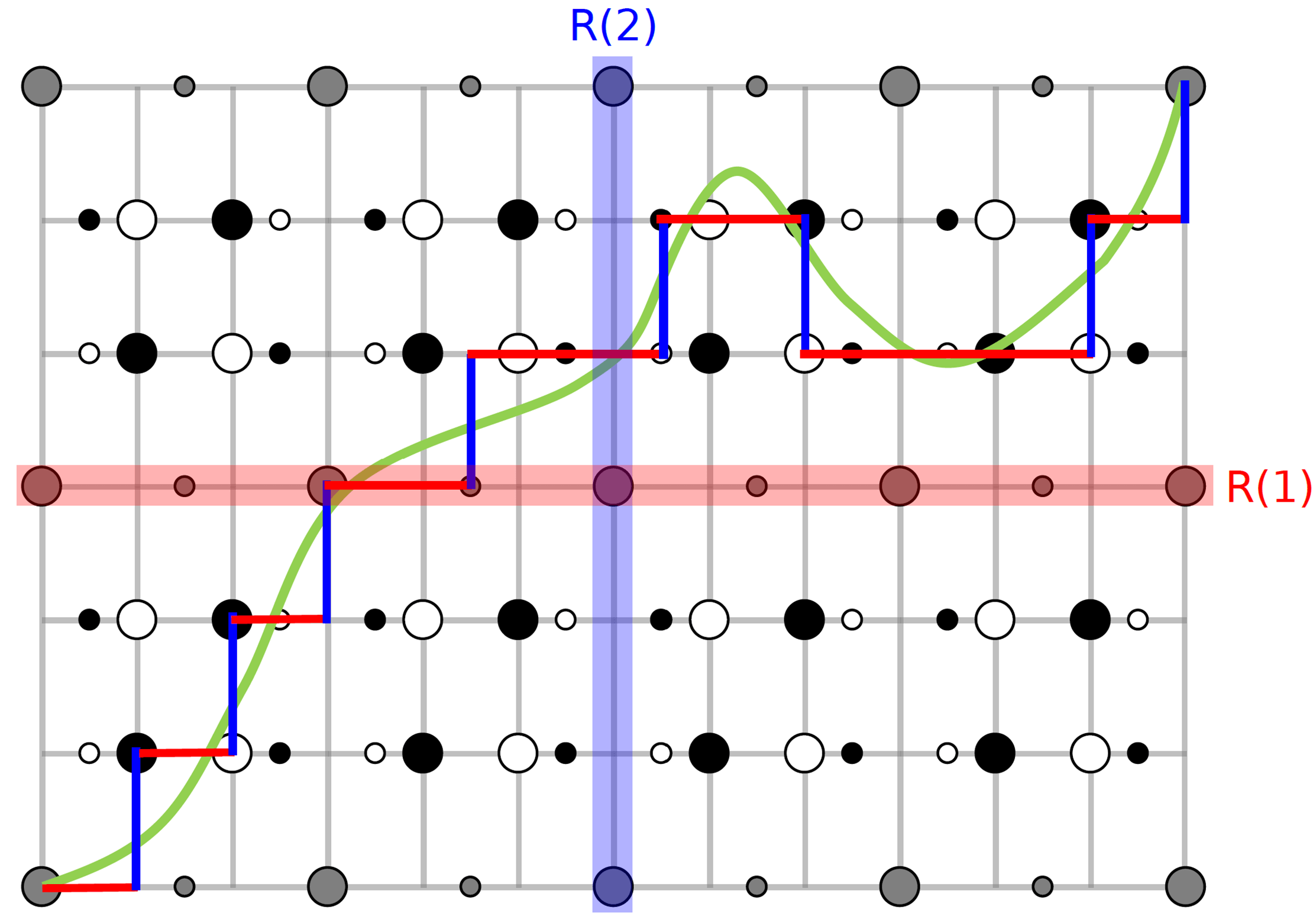}\hspace{-1.78em}%
\caption{Dichromatic pattern and DSC lattice for a $\Sigma$ 3 [110] tilt GB in an FCC crystal.
The red and blue thick lines represent the highly-coherent reference interfaces, i.e., R(1) and R(2) interfaces.
The green line is a continuously-curved GB, while the red and blue thin lines are the disconnections describing the continuously-curved GB.}
\label{fig_SM-dichromatic}
\end{figure}

Our continuum-scale GB profile can be related to the underlying bicrystallography; see the displacement-shift-complete (DSC) lattice in Fig.~\ref{fig_SM-dichromatic}.
However, more than two references must be considered in cases like the [100] tilt GBs and [111] tilt GBs in FCC crystals. In the presence of multiple references, they are not all aligned with the $\mathbf{e}_1$ and $\mathbf{e}_2$ directions. The reference interfaces, instead, are aligned along $\lbrace\mathbf{e}^{(k)}\rbrace$-axis, where $k \in [1, n]$ and n is the number of reference interfaces.
The inclination angle of each reference interface is $\phi^{(k)}$ = arccos($\mathbf{e}^{(k)}\cdot\mathbf{e}_1$), while the inclination angle of the local tangent vector $\mathbf{l}$ is $\phi(s)$ = arccos($\mathbf{l}(s)\cdot\mathbf{e}_1$). As recently exploited in the extension of the single-mode disconnection model to more-than-two references system \cite{qiu_interface_2023}, if the inclination angle $\phi$ of an interface segment lies in the range [$\phi^{(k)}$, $\phi^{(k+1)}$], reference interfaces R($k$) and R($k+1$) will be selected as the reference interfaces for this segment.

In non-orthogonal reference coordination systems, vectors defined in Cartesian coordination can also be represented in ($\mathbf{e}^{(k)}, \mathbf{e}^{(k+1)}$) as:
\begin{equation}\label{eq5}
	\left(\begin{array}{ccc}
	x_1 \\
	x_2
	\end{array}\right)
	=
	\left(\begin{array}{ccc}
	\mathbf{e}_1\cdot\mathbf{e}^{(k)}   &  \mathbf{e}_1\cdot\mathbf{e}^{(k+1)} \\
	\mathbf{e}_2\cdot\mathbf{e}^{(k)}   &  \mathbf{e}_2\cdot\mathbf{e}^{(k+1)}
	\end{array}\right)
	\left(\begin{array}{ccc}
	x^{(k)} \\
	x^{(k+1)}
	\end{array}\right) .
\end{equation}
The value of $k$ for each vector $\mathbf{x}$ depends on the inclination angle of its tangent value $\phi(x) = $ arcsin(${x_2}/{x_1}$).

We define $\rho_m^{(k)}(s){\rm d}L$ as the number of $m^{\rm th}$ mode R($k$) disconnections in the segment d$L$.
$\rho_m^{(k)} >$ 0 if, when we go along the $\mathbf{e}^{(k)}$ direction, R($k$) step leads to increment d$x_2 >$ 0 along the $\mathbf{l}$ direction.
Similarly, $\rho_m^{(k+1)} >$ 0 if, when we go along the $\mathbf{e}^{(k+1)}$ direction, R($k$+1) step leads to increment d$x_1 <$ 0 along the $\mathbf{l}$ direction.
At a point on an interface/curve in a 2D space, the density of mode-$m$ disconnections on the R($k$) interface is $\rho_m^{(k)}$.
We define the step-height vector: $\mathbf{h}^{(k)} \equiv (h_1^{(k)}, h_2^{(k)}, \cdots)^{\text{T}}$ and the R($k$)-disconnection density vector: $\boldsymbol{\rho}^{(k)} \equiv (\rho_1^{(k)}, \rho_2^{(k)}, \cdots)^{\text{T}}$.
$\mathbf{h}^{(k)}$ is a constant while $\boldsymbol{\rho}^{(k)} = \boldsymbol{\rho}^{(k)}(s)$ is a function of the parameter $s$.

R($k$) and R($k$+1) disconnections contribute to an increment of $x^{(k)}$ and $x^{(k+1)}$ by
\begin{subequations}\label{eq6}
	\begin{align}
	&\mathbf{h}^{(k)}\cdot \boldsymbol{\rho}^{(k)}{\rm d}L
	=
	{\rm d}x^{(k+1)}   \\
	&\mathbf{h}^{(k+1)}\cdot \boldsymbol{\rho}^{(k+1)}{\rm d}L
	=
	-{\rm d}x^{(k)}.
	\end{align}
\end{subequations}
The local tangent vector $\mathbf{l}$ in Cartesian coordination can be expressed in ($\mathbf{e}^{(k)}, \mathbf{e}^{(k+1)}$) as:
\begin{equation}\label{eq8}
	\left(\begin{array}{ccc}
	-\mathbf{h}^{(k+1)}\cdot\boldsymbol{\rho}^{(k+1)} \\
	\mathbf{h}^{(k)}\cdot\boldsymbol{\rho}^{(k)}
	\end{array}\right)
	= \mathbf{T}^{-1}\left(\begin{array}{ccc}
	l_1 \\
	l_2
	\end{array}\right),
\end{equation}
where $\mathbf{T}$ is the transformation tensor between the multi-reference coordinates and the Cartesian coordinate system:
%\begin{widetext}
\begin{equation}
\begin{split}
	 &\mathbf{T} = \left(\begin{array}{ccc}
	\mathbf{e}_1\cdot\mathbf{e}^{(k)}   &  \mathbf{e}_1\cdot\mathbf{e}^{(k+1)} \\
	\mathbf{e}_2\cdot\mathbf{e}^{(k)}   &  \mathbf{e}_2\cdot\mathbf{e}^{(k+1)}
	\end{array}\right)
	= \left(\begin{array}{ccc}
	\cos\phi^{(k)}   &  \cos\phi^{(k+1)} \\
	\sin\phi^{(k)}   &  \sin \phi^{(k+1)}
	\end{array}\right)
, \\
	 &\mathbf{T}^{-1}
	=
	\frac{1}{\sin(\phi^{(k+1)} - \phi^{(k)})} \left(\begin{array}{ccc}
	-\sin\phi^{(k+1)} &
	\cos\phi^{(k+1)} \\
	-\sin\phi^{(k)} &
	\cos\phi^{(k)}
	\end{array}\right).
	\end{split}
\end{equation}
%\end{widetext}
Since $\mathbf{l}$ is a unit vector, there is a geometric constraint on the densities as $l_1^2 + l_2^2 = 1$. An additional condition should be then fulfilled, which is here expressed as
\begin{equation}\label{constraintoneref}
\begin{aligned}
\mathcal{G} = &\left[\mathbf{h}^{(k+1)} \cdot \boldsymbol{\rho}^{(k+1)}\right]^2 + \left[\mathbf{h}^{(k)}  \cdot \boldsymbol{\rho}^{(k)}\right]^2  -2\mathbf{h}^{(k)}  \boldsymbol{\rho}^{(k)} \\& \cdot \mathbf{h}^{(k+1)}  \boldsymbol{\rho}^{(k+1)}
\cos (\phi^{(k+1)} - \phi^{(k)}) - 1 = 0.
\end{aligned}
\end{equation}

We note that when only one disconnection mode is activated (i.e., $m = 1$, Eq.~\eqref{constraintoneref} is naturally fulfilled.
Then, equation of motion of single-mode disconnection-mediated GB migration is simple and completely and thoroughly provided in previous references \cite{han2022disconnection, qiu_interface_2023}.
Here, we only give the details of the multiple disconnection mode cases (i.e., $m > 1$ and Eq.~\eqref{constraintoneref} is not naturally fulfilled).

\subsection{Dynamics equation for the multiple disconnection mode model}
Here, we only provide the dynamics equations and skip the derivations since they can be found in Refs.~\cite{han2022disconnection, qiu_interface_2023,qiu_disconnection_2024}.

Generalized equations of motions considering capillarity force, elastic interaction, and free energy density jump to describe GB kinetics based on a multi-mode, multi-reference disconnection model are
%\begin{tiny}
\begin{equation}\label{EOM_x}
\dot{\mathbf{x}}
=
\mathbf{M}_x \left[ \mathbf{T} \sum_m \left(\Gamma\mathbf{l}^\prime_m  - \boldsymbol{\Lambda}_m \boldsymbol{\tau}\right) + \psi \mathbf{n}\right],
\end{equation}
%\end{tiny}
where $\mathbf{M}_x$ is the diagonal mobility tensor for R(k) disconnection gliding as $\mathbf{M}_x= \mathrm{diag}(M_{x, 1}^{(k)}, M_{x, 2}^{(k)}, \cdots, \\ M_{x, 1}^{(k+1)}, M_{x, 2}^{(k+1)}, \cdots)$, $\Gamma$ is the GB stiffness, $\psi$ represents the synthetic driving force acting on the step character, $\hat{\mathbf{l}}_m^\prime$ represents the local curvature expressed by the first derivative of the density of the disconnection mode $m$, $\boldsymbol{\Lambda}_m$ refers to the term with shear-coupling factor of disconnection mode $m$ as
\begin{equation}
\boldsymbol{\Lambda}_m = \left(\begin{array}{cc}
B_m^{(k)} h_m^{(k)}\rho_m^{(k)} &  \frac{\sum_n h_n^{(k)}\rho_n^{(k)}}{\sum_n h_n^{(k+1)}\rho_n^{(k+1)}} \left|h_m^{(k)} \rho_m^{(k)}\right| \cdot \sum_n B_n^{(k+1)} h_n^{(k+1)} \rho_n^{(k+1)} \\
\frac{\sum_n h_n^{(k+1)}\rho_n^{(k+1)}}{\sum_n h_n^{(k)}\rho_n^{(k)}} \left|h_m^{(k+1)} \rho_m^{(k+1)}\right| \cdot \sum_n B_n^{(k)} h_n^{(k)} \rho_n^{(k)} &  B_m^{(k+1)} h_m^{(k+1)}\rho_m^{(k+1)}
\end{array}\right).
\end{equation}

Here, we found that, to solve the equation of motion Eq.~\eqref{EOM_x}, we need to know the temporal value of $\rho_m^{(k)}$ since the disconnection densities are not bijectively related to the local unit tangent, see Eq.~\eqref{constraintoneref}.
The equation of motion for $\rho_m^{(k)}$s is
\begin{small}
\begin{equation}\label{EOM_rho}
\begin{gathered}
\left(\begin{array}{c}
\dot{\boldsymbol{\rho}}^{(k)} \\ \dot{\boldsymbol{\rho}}^{(k+1)}
\end{array}\right)
=\\
\mathbf{M}_\rho
\left(\begin{array}{cc}
-(2\lambda|\mathbf{l}|^{-1})\mathbf{H}^{(k,k)} + \boldsymbol{\mathcal{B}}^{(k,k)}  &
\gamma_{,\phi}(1 + \psi \mathbf{H}^{(k,k+1)}/\gamma^{(k+1)})\mathbf{H}^{(k,k+1)} + \boldsymbol{\mathcal{B}}^{(k,k+1)} \\
-\gamma_{,\phi}(1 + \psi \mathbf{H}^{(k+1,k)}/\gamma^{(k)} )\mathbf{H}^{(k+1,k)} + \boldsymbol{\mathcal{B}}^{(k+1,k)} &
-(2\lambda|\mathbf{l}|^{-1}) \mathbf{H}^{(k+1,k+1)} + \boldsymbol{\mathcal{B}}^{(k+1,k+1)}
\end{array}\right)
\left(\begin{array}{c}
\boldsymbol{\rho}^{(k)} \\ \boldsymbol{\rho}^{(k+1)}
\end{array}\right),
\end{gathered}
\end{equation}
\end{small}
where $\mathbf{M}_\rho^{(k)}$ is the diagonal mobility tensor for R(k) disconnection density evolution as $\mathbf{M}^{(k)} = \mathrm{diag} (M_{1, \rho}^{(k)}, M_{2, \rho}^{(k)}, \cdots)$, $\mathbf{H}^{(k, l)} = \mathbf{h}^{(k)} \otimes \mathbf{h}^{(l)}$ and  $\gamma_{,\phi}$ is the first derivative of GB energy with respect to the inclination angle.
$\gamma^{(k)}$ and $\gamma^{(k+1)}$ are the GB energies of the R(k) and R(k+1) interfaces.
The nonlocal operator $\boldsymbol{\mathcal{B}}^{(kl)}$ is related to the driving force caused by self-stress among disconnections as
\begin{equation}\label{hatBklmn}
\mathcal{B}_{mn}^{(kl)} g(s)
\equiv
 \int \rmd s_0
(\mathbf{b}_m^{(k)} \times \boldsymbol{\xi})
\cdot
\left\{
\frac{\mu(1+\delta_{km,ln})|\mathbf{l}|}{2\pi(1-\nu)}
\left[
\ln\frac{r(s,s_0)}{r_0} \mathbf{I}
+ \hat{\mathbf{r}}(s,s_0) \otimes \hat{\mathbf{r}}(s,s_0)
\right]
\right\}
(\mathbf{b}_n^{(l)} \times \boldsymbol{\xi}) \cdot g(s_0), ~
\forall \ g(s),
\end{equation}
where $\delta_{km, ln}$ is the delta function equal to 1 when $k = l$ and $m = n$.
The Lagrange multiplier $\lambda$ is also evolving to dynamically fulfill the geometric constraint with the evolving $\rho_m^{(k)}$, Eq.~\eqref{constraintoneref}.
The kinetic equation, similar to the dynamic optimization problem \cite{kameswaran2006simultaneous}, is written as
\begin{equation}\label{lambda}
\begin{split}
\dot{\lambda} = \frac{M_\lambda}{\vert \mathbf{l}\vert} \bigg\{& \left[\mathbf{h}^{(k+1)} \cdot \boldsymbol{\rho}^{(k+1)}\right]^2 + \left[\mathbf{h}^{(k)}  \cdot \boldsymbol{\rho}^{(k)}\right]^2 -2\mathbf{h}^{(k)}  \boldsymbol{\rho}^{(k)}  \\
&  \cdot \mathbf{h}^{(k+1)}  \boldsymbol{\rho}^{(k+1)} \cos (\phi^{(k+1)} - \phi^{(k)}) - 1 \bigg\},
\end{split}
\end{equation}
where $M_\lambda$ is the penalty coefficient.

\subsection{Numerical implementation}
We conduct a series of numerical simulations to study  disconnection flow-mediated grain rotation via   grain shrinkage/growth of an embedded grain.
Numerical simulations are performed by solving the equations of motion (i.e, ~\eqref{EOM_x}, \eqref{EOM_rho} and \eqref{lambda}).

We employ the reduced units for the quantities as follows: $\tilde{\mathbf{x}} = \mathbf{x}/\alpha$, $\tilde{\kappa} = \kappa\alpha$, $\Delta\tilde{t} = \Delta t M_{x, 1}^{(1)} \gamma^{(1)}/\alpha^2$, $\tilde{M}_{x, m}^{(k)} = M_{x, m}^{(k)}/M_{x, 1}^{(1)}$, $\tilde{M}_{\rho, m}^{(k)} = M_{\rho, m}^{(k)}/M_{x, 1}^{(1)}$, $\tilde{\tau} = \tau\alpha/\gamma^{(1)}$,  $\tilde{\psi} = \psi\alpha/\gamma^{(1)}$, $\tilde{T} = k_B T/\gamma^{(1)}$, $\tilde{Q} =Q/(\gamma^{(1)}\alpha)$, $\tilde{\gamma} =\gamma/\gamma^{(1)}$, where $\alpha$ is the DSC lattice parameter.

The interface curve $\tilde{\mathbf{x}}(s)$ is discretized into N segments with $\left\lbrace\tilde{\mathbf{x}}(i)\right\rbrace$ with $n$ = 0, 1, ..., $N -1$.
Derivatives $\tilde{\mathbf{x}}'$ and $\tilde{\mathbf{x}}''$ are computed using a three-point stencil $\left\lbrace\tilde{\mathbf{x}}(i-1), {\mathbf{x}}(i), {\mathbf{x}}(i+1)\right\rbrace$.
Centred finite differences with periodic boundary conditions ($\tilde{\mathbf{x}}(N) = \tilde{\mathbf{x}}(0)$), which are expressed as:
\begin{align}\label{eq19}
	\left(\frac{{\rm d}\tilde{\mathbf{x}}}{{\rm d}s}\right)_{\bar s}
	&=
	\frac{\tilde{\mathbf{x}}(\bar s+\Delta s)-\tilde{\mathbf{x}}(\bar s-\Delta s)}{2\Delta s} + \mathcal{O}[(\Delta s)],\\
	\left(\frac{{\rm d}^2\tilde{\mathbf{x}}}{{\rm d}s^2}\right)_{\bar s}
	&=
	\frac{\tilde{\mathbf{x}}(\bar s+\Delta s)+\tilde{\mathbf{x}}(\bar s-\Delta s)-2\tilde{\mathbf{x}}(\bar s)}{(\Delta s)^2} + \mathcal{O}[(\Delta s)^2].
\end{align}
or as a finite difference,
\begin{align}\label{eq21_22}
	\mathbf{l}(i)
	&=
	\tilde{\mathbf{x}}'(s)
	=
	[\tilde{\mathbf{x}}(i+1)-\tilde{\mathbf{x}}(i-1)]/2 \\
	\tilde{\mathbf{x}}''(s)
	&=
	\tilde{\mathbf{x}}(i+1)+\tilde{\mathbf{x}}(i-1)-2\tilde{\mathbf{x}}(i)
\end{align}

\subsection{Materials parameters}
The shape of the initial embedded grain in all simulations is a circle of radius 100.
The applied synthetic driving force in the two-mode cases is set as -0.02, while that for the single-mode case is set as -0.04.
Lattice constant $\alpha$ is 3.52 \text{\AA}, and temperature $T$ = 1000 K are chosen for all cases in this section.
The details of the disconnection Burgers vectors and step heights of all these three CSL GBs are given in Table.~\ref{tab:R4-parameters2}.
As for [100] tilt GBs in FCC crystals, $\gamma^{(1)} = \gamma^{(3)},~ \gamma^{(2)} = \gamma^{(4)}$, and $M_{z, m}^{(1)} = M_{z, m}^{(3)},~ M_{z, m}^{(2)} = M_{z, m}^{(4)}$ because of the crystallographic symmetry.
The remaining materials parameters in this case are provided in Table~\ref{tab:R4-parameters1}.
We here only provide the value of $M_{x, 1}^{(1)} = M_{x, 1}^{(3)}$ and the ratio $M_{\rho, 1}^{(1)}/M_{x, 1}^{(1)}$. The other mobilities $M_{z, m}^{(k)}$ can be calculated as
\begin{align}
&\frac{M_{x, m}^{(k)}}{M_{x,1}^{(1)}} = \exp\left( - \frac{Q_m^{(k)} - Q_1^{(1)}}{k_{\rm B} T} \right) \\
&\frac{M_{\rho, m}^{(k)}}{M_{\rho,1}^{(1)}} = \exp\left( - \frac{Q_m^{(k)} - Q_1^{(1)}}{k_{\rm B} T} \right)
\end{align}
Here, we scaled the reduced mobility as $\min\{M_{x,m}^{k}\} = 1$.

\begin{table*}[htpb!]
\caption{\label{tab:R4-parameters2} Crystallographic Parameters used in the simulations of [100] tilt GBs in Ni.}
\renewcommand{\arraystretch}{1.5}
%\begin{ruledtabular}
\begin{tabular}{c cccc cccccccc}
%%%%%%%%%%%%%%%%%%%%%%%%%%%%%%%%%
\hline
{CSL} &
\multicolumn{4}{c}{Burgers vector} &
\multicolumn{8}{c}{Step height}\\
 \cline{2-5} \cline{6-13}
%%%%%%%%%%%%%%%%%%%%%%%%%%%%%%%%%
 &
$b^{(1)}$ & $b^{(2)}$  & $b^{(3)}$  & $b^{(4)}$    &
$h^{(1)}_1$ & $h^{(1)}_2$  & $h^{(2)}_1$  & $h^{(2)}_2$  & $h^{(3)}_1$ & $h^{(3)}_2$  & $h^{(4)}_1$  & $h^{(4)}_2$  \\
\hline
%\cline{1-1} \cline{2-5} \cline{6-9} \cline{10-17}
%%%%%%%%%%%%%%%%%%%%%%%%%%%%%%%%%
{\shortstack{$\Sigma 5$}} &
$\sqrt{5}\alpha/5$ & $\sqrt{2}b^{(1)}$ & $b^{(1)}$  & $\sqrt{2}b^{(1)}$ &
-$b^{(3)}$  &  $\mathbf{1.5b^{(3)}}$ & $\mathbf{-b^{(4)}}$  &  1.5$b^{(4)}$ & $b^{(1)}$  &  $\mathbf{-1.5b^{(1)}}$ & $\mathbf{b^{(2)}}$  &  -1.5$b^{(2)}$  \\
\hline
%%%%%%%%%%%%%%%%%%%%%%%%%%%%%%%%%
{\shortstack{$\Sigma 17$}} &
%1 & $\sqrt{2}$ &1 & $\sqrt{2}$  &
$\sqrt{17}\alpha/17$ & $\sqrt{2}b^{(1)}$ & $b^{(1)}$  & $\sqrt{2}b^{(1)}$ &
$\mathbf{-2b^{(3)}}$  &  6.5$b^{(3)}$ & $\mathbf{-2b^{(4)}}$  &  6.5$b^{(4)}$ & $\mathbf{2b^{(1)}}$  &  -6.5$b^{(1)}$ & $\mathbf{2b^{(2)}}$  &  -6.5$b^{(2)}$  \\
\hline
%%%%%%%%%%%%%%%%%%%%%%%%%%%%%%%%%
{\shortstack{$\Sigma 29$}} &
%1 & $\sqrt{2}$ &1 & $\sqrt{2}$  &
$\sqrt{29}\alpha/29$ & $\sqrt{2}b^{(1)}$ & $b^{(1)}$  & $\sqrt{2}b^{(1)}$ &
6$b^{(3)}$  &  $\mathbf{-8.5b^{(3)}}$ & $\mathbf{6b^{(4)}}$  &  -8.5$b^{(4)}$ & -6$b^{(1)}$  &  $\mathbf{8.5b^{(1)}}$ & $\mathbf{-6b^{(2)}}$  &  8.5$b^{(2)}$  \\
\hline\\
\end{tabular}
%\end{ruledtabular}
\end{table*}

\begin{table*}[htpb!]
\centering
\caption{\label{tab:R4-parameters1} Materials parameters for [100] tilt GBs in Ni, n = 1 and 3.}
\renewcommand{\arraystretch}{1}
%\begin{ruledtabular}
\begin{tabular}{cccccccc}
%%%%%%%%%%%%%%%%%%%%%%%%%%%%%%%%%
\hline
{CSL} &
\multicolumn{2}{c}{GB energy (J/m$^2$)} &
\multicolumn{4}{c}{Gliding energy barrier $Q$ (eV/nm)}  \\
\cline{2-3} \cline{4-7}
%%%%%%%%%%%%%%%%%%%%%%%%%%%%%%%%%
 & $\gamma^{(n)}$  & $\gamma^{(n+1)}$ & $Q_1^{(n)}$  & $Q_2^{(n)}$ & $Q_1^{(n+1)}$  & $Q_2^{(n+1)}$   \\
\hline
%%%%%%%%%%%%%%%%%%%%%%%%%%%%%%%%%
$\Sigma 5$ & 0.90 & 0.95 & 1.0 & $\mathbf{0.2}$ & $\mathbf{0.1}$ & 1.2  \\
$\Sigma 17$ & 0.91 & 0.85 & $\mathbf{1.0}$ & 1.1 & $\mathbf{0.25}$ & 1.2 \\
$\Sigma 29$ & 0.98 & 0.95 & 1.8 & $\mathbf{1.4}$ & $\mathbf{2.0}$ & 2.1\\
\hline
\end{tabular}
%\end{ruledtabular}
\end{table*}

In the case of single disconnection mode, we only choose the disconnection mode with lower activation energy between the two modes provided in Tab.~\ref{tab:R4-parameters2} (see the \emph{bold} text). 

%\end{frontmatter}

\setcounter{figure}{0}


\newpage
\begin{thebibliography}{10}
\expandafter\ifx\csname url\endcsname\relax
  \def\url#1{\texttt{#1}}\fi
\expandafter\ifx\csname urlprefix\endcsname\relax\def\urlprefix{URL }\fi
\expandafter\ifx\csname href\endcsname\relax
  \def\href#1#2{#2} \def\path#1{#1}\fi

\bibitem{rohrer_grain_2011}
G.~S. Rohrer, \href{https://doi.org/10.1007/s10853-011-5677-3}{Grain boundary energy anisotropy: a review}, Journal of Materials Science 46~(18) (2011) 5881--5895.
\newblock \href {http://dx.doi.org/10.1007/s10853-011-5677-3} {\path{doi:10.1007/s10853-011-5677-3}}.
\newline\urlprefix\url{https://doi.org/10.1007/s10853-011-5677-3}

\bibitem{saylor_habits_2004}
D.~M. Saylor, B.~El~Dasher, Y.~Pang, H.~M. Miller, P.~Wynblatt, A.~D. Rollett, G.~S. Rohrer, \href{https://onlinelibrary.wiley.com/doi/abs/10.1111/j.1551-2916.2004.00724.x}{Habits of {Grains} in {Dense} {Polycrystalline} {Solids}}, Journal of the American Ceramic Society 87~(4) (2004) 724--726, \_eprint: https://onlinelibrary.wiley.com/doi/pdf/10.1111/j.1551-2916.2004.00724.x.
\newblock \href {http://dx.doi.org/10.1111/j.1551-2916.2004.00724.x} {\path{doi:10.1111/j.1551-2916.2004.00724.x}}.
\newline\urlprefix\url{https://onlinelibrary.wiley.com/doi/abs/10.1111/j.1551-2916.2004.00724.x}

\bibitem{goodhew_low_1978}
P.~J. Goodhew, T.~Y. Tan, R.~W. Balluffi, \href{https://www.sciencedirect.com/science/article/pii/0001616078901086}{Low energy planes for tilt grain boundaries in gold}, Acta Metallurgica 26~(4) (1978) 557--567.
\newblock \href {http://dx.doi.org/10.1016/0001-6160(78)90108-6} {\path{doi:10.1016/0001-6160(78)90108-6}}.
\newline\urlprefix\url{https://www.sciencedirect.com/science/article/pii/0001616078901086}

\bibitem{muschik_energetic_1993}
T.~Muschik, W.~Laub, U.~Wolf, M.~W. Finnis, W.~Gust, \href{https://www.sciencedirect.com/science/article/pii/0956715193903867}{Energetic and kinetic aspects of the faceting transformation of a Σ3 grain boundary in {Cu}}, Acta Metallurgica et Materialia 41~(7) (1993) 2163--2171.
\newblock \href {http://dx.doi.org/10.1016/0956-7151(93)90386-7} {\path{doi:10.1016/0956-7151(93)90386-7}}.
\newline\urlprefix\url{https://www.sciencedirect.com/science/article/pii/0956715193903867}

\bibitem{barg_faceting_1995}
A.~I. Barg, E.~Rabkin, W.~Gust, \href{https://www.sciencedirect.com/science/article/pii/095671519500094C}{Faceting transformation and energy of a Σ3 grain boundary in silver}, Acta Metallurgica et Materialia 43~(11) (1995) 4067--4074.
\newblock \href {http://dx.doi.org/10.1016/0956-7151(95)00094-C} {\path{doi:10.1016/0956-7151(95)00094-C}}.
\newline\urlprefix\url{https://www.sciencedirect.com/science/article/pii/095671519500094C}

\bibitem{dobrovolski2024facet}
B.~Dobrovolski, C.~B. Owens, G.~L. Hart, E.~R. Homer, B.~Runnels, Facet and energy predictions in grain boundaries: Lattice matching and molecular dynamics, Acta Materialia 274 (2024) 119962.

\bibitem{merkle_low-energy_1992}
K.~L. Merkle, D.~Wolf, \href{https://doi.org/10.1080/01418619208201536}{Low-energy configurations of symmetric and asymmetric tilt grain boundaries}, Philosophical Magazine A 65~(2) (1992) 513--530, publisher: Taylor \& Francis \_eprint: https://doi.org/10.1080/01418619208201536.
\newblock \href {http://dx.doi.org/10.1080/01418619208201536} {\path{doi:10.1080/01418619208201536}}.
\newline\urlprefix\url{https://doi.org/10.1080/01418619208201536}

\bibitem{xu_energy_2023}
Z.~Xu, C.~M. Hefferan, S.~F. Li, J.~Lind, R.~M. Suter, F.~Abdeljawad, G.~S. Rohrer, \href{https://www.sciencedirect.com/science/article/pii/S135964622300129X}{Energy dissipation by grain boundary replacement during grain growth}, Scripta Materialia 230 (2023) 115405.
\newblock \href {http://dx.doi.org/10.1016/j.scriptamat.2023.115405} {\path{doi:10.1016/j.scriptamat.2023.115405}}.
\newline\urlprefix\url{https://www.sciencedirect.com/science/article/pii/S135964622300129X}

\bibitem{xu_grain_2024}
Z.~Xu, Y.-F. Shen, S.~K. Naghibzadeh, X.~Peng, V.~Muralikrishnan, S.~Maddali, D.~Menasche, A.~R. Krause, K.~Dayal, R.~M. Suter, G.~S. Rohrer, \href{https://www.sciencedirect.com/science/article/pii/S1359645423008704}{Grain boundary migration in polycrystalline α-{Fe}}, Acta Materialia 264 (2024) 119541.
\newblock \href {http://dx.doi.org/10.1016/j.actamat.2023.119541} {\path{doi:10.1016/j.actamat.2023.119541}}.
\newline\urlprefix\url{https://www.sciencedirect.com/science/article/pii/S1359645423008704}

\bibitem{wulff_xxv_1901}
G.~Wulff, \href{https://www.degruyter.com/document/doi/10.1524/zkri.1901.34.1.449/html}{{XXV}. {Zur} {Frage} der {Geschwindigkeit} des {Wachsthums} und der {Auflösung} der {Krystallflächen}}, Zeitschrift für Kristallographie - Crystalline Materials 34~(1-6) (1901) 449--530, publisher: De Gruyter (O).
\newblock \href {http://dx.doi.org/10.1524/zkri.1901.34.1.449} {\path{doi:10.1524/zkri.1901.34.1.449}}.
\newline\urlprefix\url{https://www.degruyter.com/document/doi/10.1524/zkri.1901.34.1.449/html}

\bibitem{herring_theorems_1951}
C.~Herring, \href{https://link.aps.org/doi/10.1103/PhysRev.82.87}{Some {Theorems} on the {Free} {Energies} of {Crystal} {Surfaces}}, Physical Review 82~(1) (1951) 87--93, publisher: American Physical Society.
\newblock \href {http://dx.doi.org/10.1103/PhysRev.82.87} {\path{doi:10.1103/PhysRev.82.87}}.
\newline\urlprefix\url{https://link.aps.org/doi/10.1103/PhysRev.82.87}

\bibitem{sekerka_equilibrium_2005}
R.~F. Sekerka, \href{https://onlinelibrary.wiley.com/doi/abs/10.1002/crat.200410342}{Equilibrium and growth shapes of crystals: how do they differ and why should we care?}, Crystal Research and Technology 40~(4-5) (2005) 291--306, \_eprint: https://onlinelibrary.wiley.com/doi/pdf/10.1002/crat.200410342.
\newblock \href {http://dx.doi.org/10.1002/crat.200410342} {\path{doi:10.1002/crat.200410342}}.
\newline\urlprefix\url{https://onlinelibrary.wiley.com/doi/abs/10.1002/crat.200410342}

\bibitem{rheinheimer_equilibrium_2020}
W.~Rheinheimer, J.~E. Blendell, C.~A. Handwerker, \href{https://www.sciencedirect.com/science/article/pii/S1359645420302512}{Equilibrium and kinetic shapes of grains in polycrystals}, Acta Materialia 191 (2020) 101--110.
\newblock \href {http://dx.doi.org/10.1016/j.actamat.2020.03.055} {\path{doi:10.1016/j.actamat.2020.03.055}}.
\newline\urlprefix\url{https://www.sciencedirect.com/science/article/pii/S1359645420302512}

\bibitem{hanyu_stress-induced_2005}
S.~Hanyu, H.~Nishimura, K.~Matsunaga, T.~Yamamoto, Y.~Ikuhara, A.~M. Glaeser, \href{https://doi.org/10.1007/s10853-005-2675-3}{Stress-induced facet coarsening in a σ\$\$7{\textbackslash}\{4{\textbackslash}bar\{5\}10{\textbackslash}\}\$\$symmetrical tilt grain boundary in an alumina bicrystal}, Journal of Materials Science 40~(12) (2005) 3137--3142.
\newblock \href {http://dx.doi.org/10.1007/s10853-005-2675-3} {\path{doi:10.1007/s10853-005-2675-3}}.
\newline\urlprefix\url{https://doi.org/10.1007/s10853-005-2675-3}

\bibitem{qiu_interface_2023}
C.~Qiu, M.~Salvalaglio, D.~J. Srolovitz, J.~Han, \href{https://www.sciencedirect.com/science/article/pii/S1359645423002112}{Interface faceting–defaceting mediated by disconnections}, Acta Materialia 251 (2023) 118880.
\newblock \href {http://dx.doi.org/10.1016/j.actamat.2023.118880} {\path{doi:10.1016/j.actamat.2023.118880}}.
\newline\urlprefix\url{https://www.sciencedirect.com/science/article/pii/S1359645423002112}

\bibitem{qiu_disconnection_2024}
C.~Qiu, M.~Salvalaglio, D.~J. Srolovitz, J.~Han, \href{https://www.pnas.org/doi/10.1073/pnas.2310302121}{Disconnection flow–mediated grain rotation}, Proceedings of the National Academy of Sciences 121~(1) (2024) e2310302121, publisher: Proceedings of the National Academy of Sciences.
\newblock \href {http://dx.doi.org/10.1073/pnas.2310302121} {\path{doi:10.1073/pnas.2310302121}}.
\newline\urlprefix\url{https://www.pnas.org/doi/10.1073/pnas.2310302121}

\bibitem{schratt_grain_2021}
A.~A. Schratt, I.~Steinbach, V.~Mohles, \href{https://www.sciencedirect.com/science/article/pii/S0927025621001099}{Grain boundary energy landscape from the shape analysis of synthetically stabilized embedded grains}, Computational Materials Science 193 (2021) 110384.
\newblock \href {http://dx.doi.org/10.1016/j.commatsci.2021.110384} {\path{doi:10.1016/j.commatsci.2021.110384}}.
\newline\urlprefix\url{https://www.sciencedirect.com/science/article/pii/S0927025621001099}

\bibitem{plimpton_fast_1995}
S.~Plimpton, \href{https://www.sciencedirect.com/science/article/pii/S002199918571039X}{Fast {Parallel} {Algorithms} for {Short}-{Range} {Molecular} {Dynamics}}, Journal of Computational Physics 117~(1) (1995) 1--19.
\newblock \href {http://dx.doi.org/10.1006/jcph.1995.1039} {\path{doi:10.1006/jcph.1995.1039}}.
\newline\urlprefix\url{https://www.sciencedirect.com/science/article/pii/S002199918571039X}

\bibitem{thompson_lammps_2022}
A.~P. Thompson, H.~M. Aktulga, R.~Berger, D.~S. Bolintineanu, W.~M. Brown, P.~S. Crozier, P.~J. in~'t Veld, A.~Kohlmeyer, S.~G. Moore, T.~D. Nguyen, R.~Shan, M.~J. Stevens, J.~Tranchida, C.~Trott, S.~J. Plimpton, \href{https://www.sciencedirect.com/science/article/pii/S0010465521002836}{{LAMMPS} - a flexible simulation tool for particle-based materials modeling at the atomic, meso, and continuum scales}, Computer Physics Communications 271 (2022) 108171.
\newblock \href {http://dx.doi.org/10.1016/j.cpc.2021.108171} {\path{doi:10.1016/j.cpc.2021.108171}}.
\newline\urlprefix\url{https://www.sciencedirect.com/science/article/pii/S0010465521002836}

\bibitem{foiles_computation_2006}
S.~M. Foiles, J.~J. Hoyt, \href{https://www.sciencedirect.com/science/article/pii/S1359645406002333}{Computation of grain boundary stiffness and mobility from boundary fluctuations}, Acta Materialia 54~(12) (2006) 3351--3357.
\newblock \href {http://dx.doi.org/10.1016/j.actamat.2006.03.037} {\path{doi:10.1016/j.actamat.2006.03.037}}.
\newline\urlprefix\url{https://www.sciencedirect.com/science/article/pii/S1359645406002333}

\bibitem{hestenes_methods_1952}
M.~Hestenes, E.~Stiefel, Methods of conjugate gradients for solving linear systems, Journal of research of the National Bureau of Standards 49~(6).
\newblock \href {http://dx.doi.org/10.6028/JRES.049.044} {\path{doi:10.6028/JRES.049.044}}.

\bibitem{trautt_grain_2012}
Z.~T. Trautt, Y.~Mishin, \href{https://www.sciencedirect.com/science/article/pii/S1359645412000432}{Grain boundary migration and grain rotation studied by molecular dynamics}, Acta Materialia 60~(5) (2012) 2407--2424.
\newblock \href {http://dx.doi.org/10.1016/j.actamat.2012.01.008} {\path{doi:10.1016/j.actamat.2012.01.008}}.
\newline\urlprefix\url{https://www.sciencedirect.com/science/article/pii/S1359645412000432}

\bibitem{qiu_variability_2023}
A.~Qiu, I.~Chesser, E.~Holm, \href{https://www.sciencedirect.com/science/article/pii/S1359645423004068}{On the variability of grain boundary mobility in the isoconfigurational ensemble}, Acta Materialia 257 (2023) 119075.
\newblock \href {http://dx.doi.org/10.1016/j.actamat.2023.119075} {\path{doi:10.1016/j.actamat.2023.119075}}.
\newline\urlprefix\url{https://www.sciencedirect.com/science/article/pii/S1359645423004068}

\bibitem{janssens_computing_2006}
K.~G.~F. Janssens, D.~Olmsted, E.~A. Holm, S.~M. Foiles, S.~J. Plimpton, P.~M. Derlet, \href{https://www.nature.com/articles/nmat1559}{Computing the mobility of grain boundaries}, Nature Materials 5~(2) (2006) 124--127, number: 2 Publisher: Nature Publishing Group.
\newblock \href {http://dx.doi.org/10.1038/nmat1559} {\path{doi:10.1038/nmat1559}}.
\newline\urlprefix\url{https://www.nature.com/articles/nmat1559}

\bibitem{schratt_efficient_2020}
A.~A. Schratt, V.~Mohles, \href{https://www.sciencedirect.com/science/article/pii/S0927025620302652}{Efficient calculation of the {ECO} driving force for atomistic simulations of grain boundary motion}, Computational Materials Science 182 (2020) 109774.
\newblock \href {http://dx.doi.org/10.1016/j.commatsci.2020.109774} {\path{doi:10.1016/j.commatsci.2020.109774}}.
\newline\urlprefix\url{https://www.sciencedirect.com/science/article/pii/S0927025620302652}

\bibitem{larsen_robust_2016}
P.~M. Larsen, S.~Schmidt, J.~Schiøtz, \href{https://dx.doi.org/10.1088/0965-0393/24/5/055007}{Robust structural identification via polyhedral template matching}, Modelling and Simulation in Materials Science and Engineering 24~(5) (2016) 055007, publisher: IOP Publishing.
\newblock \href {http://dx.doi.org/10.1088/0965-0393/24/5/055007} {\path{doi:10.1088/0965-0393/24/5/055007}}.
\newline\urlprefix\url{https://dx.doi.org/10.1088/0965-0393/24/5/055007}

\bibitem{stukowski_visualization_2009}
A.~Stukowski, \href{https://doi.org/10.1088/0965-0393/18/1/015012}{Visualization and analysis of atomistic simulation data with {OVITO}–the {Open} {Visualization} {Tool}}, Modelling and Simulation in Materials Science and Engineering 18~(1) (2009) 015012, publisher: IOP Publishing.
\newblock \href {http://dx.doi.org/10.1088/0965-0393/18/1/015012} {\path{doi:10.1088/0965-0393/18/1/015012}}.
\newline\urlprefix\url{https://doi.org/10.1088/0965-0393/18/1/015012}

\bibitem{straumal_faceting_2001}
B.~Straumal, S.~Polyakov, E.~Bischoff, W.~Gust, E.~Mittemeijer, \href{https://doi.org/10.1023/A:1015174921561}{Faceting of Σ3 and Σ9 {Grain} {Boundaries} in {Copper}}, Interface Science 9~(3) (2001) 287--292.
\newblock \href {http://dx.doi.org/10.1023/A:1015174921561} {\path{doi:10.1023/A:1015174921561}}.
\newline\urlprefix\url{https://doi.org/10.1023/A:1015174921561}

\bibitem{straumal_temperature_2006}
B.~B. Straumal, S.~A. Polyakov, E.~J. Mittemeijer, \href{https://www.sciencedirect.com/science/article/pii/S1359645405005100}{Temperature influence on the faceting of Σ3 and Σ9 grain boundaries in {Cu}}, Acta Materialia 54~(1) (2006) 167--172.
\newblock \href {http://dx.doi.org/10.1016/j.actamat.2005.08.037} {\path{doi:10.1016/j.actamat.2005.08.037}}.
\newline\urlprefix\url{https://www.sciencedirect.com/science/article/pii/S1359645405005100}

\bibitem{li_atomistic_2019}
S.~Li, L.~Yang, C.~Lai, \href{https://www.sciencedirect.com/science/article/pii/S0927025619300667}{Atomistic simulations of energies for arbitrary grain boundaries. {Part} {I}: {Model} and validation}, Computational Materials Science 161 (2019) 330--338.
\newblock \href {http://dx.doi.org/10.1016/j.commatsci.2019.02.003} {\path{doi:10.1016/j.commatsci.2019.02.003}}.
\newline\urlprefix\url{https://www.sciencedirect.com/science/article/pii/S0927025619300667}

\bibitem{lee_computation_2004}
B.-J. Lee, S.-H. Choi, \href{https://dx.doi.org/10.1088/0965-0393/12/4/005}{Computation of grain boundary energies}, Modelling and Simulation in Materials Science and Engineering 12~(4) (2004) 621.
\newblock \href {http://dx.doi.org/10.1088/0965-0393/12/4/005} {\path{doi:10.1088/0965-0393/12/4/005}}.
\newline\urlprefix\url{https://dx.doi.org/10.1088/0965-0393/12/4/005}

\bibitem{han2022disconnection}
J.~Han, D.~J. Srolovitz, M.~Salvalaglio, {Disconnection-mediated migration of interfaces in microstructures: I. continuum model}, Acta Materialia 227 (2022) 117178.
\newblock \href {http://dx.doi.org/10.1016/j.actamat.2021.117178} {\path{doi:10.1016/j.actamat.2021.117178}}.

\bibitem{chen_grain_2019}
K.~Chen, J.~Han, S.~L. Thomas, D.~J. Srolovitz, \href{https://www.sciencedirect.com/science/article/pii/S1359645419300552}{Grain boundary shear coupling is not a grain boundary property}, Acta Materialia 167 (2019) 241--247.
\newblock \href {http://dx.doi.org/10.1016/j.actamat.2019.01.040} {\path{doi:10.1016/j.actamat.2019.01.040}}.
\newline\urlprefix\url{https://www.sciencedirect.com/science/article/pii/S1359645419300552}

\bibitem{han_grain-boundary_2018}
J.~Han, S.~L. Thomas, D.~J. Srolovitz, \href{https://www.sciencedirect.com/science/article/pii/S0079642518300641}{Grain-boundary kinetics: {A} unified approach}, Progress in Materials Science 98 (2018) 386--476.
\newblock \href {http://dx.doi.org/10.1016/j.pmatsci.2018.05.004} {\path{doi:10.1016/j.pmatsci.2018.05.004}}.
\newline\urlprefix\url{https://www.sciencedirect.com/science/article/pii/S0079642518300641}

\bibitem{zhang2014lattice}
Y.~Zhang, Y.~Gao, L.~Nicola, Lattice rotation caused by wedge indentation of a single crystal: dislocation dynamics compared to crystal plasticity simulations, Journal of the Mechanics and Physics of Solids 68 (2014) 267--279.
\newblock \href {http://dx.doi.org/10.1016/j.jmps.2014.04.006} {\path{doi:10.1016/j.jmps.2014.04.006}}.

\bibitem{kareer2020scratching}
A.~Kareer, E.~Tarleton, C.~Hardie, S.~V. Hainsworth, A.~J. Wilkinson, Scratching the surface: Elastic rotations beneath nanoscratch and nanoindentation tests, Acta Materialia 200 (2020) 116--126.
\newblock \href {http://dx.doi.org/10.1016/j.actamat.2020.08.051} {\path{doi:10.1016/j.actamat.2020.08.051}}.

\bibitem{upmanyu_simultaneous_2006}
M.~Upmanyu, D.~J. Srolovitz, A.~E. Lobkovsky, J.~A. Warren, W.~C. Carter, \href{https://www.sciencedirect.com/science/article/pii/S135964540500707X}{Simultaneous grain boundary migration and grain rotation}, Acta Materialia 54~(7) (2006) 1707--1719.
\newblock \href {http://dx.doi.org/10.1016/j.actamat.2005.11.036} {\path{doi:10.1016/j.actamat.2005.11.036}}.
\newline\urlprefix\url{https://www.sciencedirect.com/science/article/pii/S135964540500707X}

\bibitem{molodov_grain_2015}
D.~A. Molodov, L.~A. Barrales-Mora, J.-E. Brandenburg, \href{https://doi.org/10.1088/1757-899x/89/1/012008}{Grain boundary motion and grain rotation in aluminum bicrystals: recent experiments and simulations}, IOP Conference Series: Materials Science and Engineering 89 (2015) 012008, publisher: IOP Publishing.
\newblock \href {http://dx.doi.org/10.1088/1757-899X/89/1/012008} {\path{doi:10.1088/1757-899X/89/1/012008}}.
\newline\urlprefix\url{https://doi.org/10.1088/1757-899x/89/1/012008}

\bibitem{french_molecular_2022}
J.~French, X.-M. Bai, \href{https://www.sciencedirect.com/science/article/pii/S002231152200232X}{Molecular dynamics studies of grain boundary mobility and anisotropy in {BCC} γ-uranium}, Journal of Nuclear Materials 565 (2022) 153744.
\newblock \href {http://dx.doi.org/10.1016/j.jnucmat.2022.153744} {\path{doi:10.1016/j.jnucmat.2022.153744}}.
\newline\urlprefix\url{https://www.sciencedirect.com/science/article/pii/S002231152200232X}

\bibitem{stukowski_automated_2012}
A.~Stukowski, V.~V. Bulatov, A.~Arsenlis, \href{https://dx.doi.org/10.1088/0965-0393/20/8/085007}{Automated identification and indexing of dislocations in crystal interfaces}, Modelling and Simulation in Materials Science and Engineering 20~(8) (2012) 085007, publisher: IOP Publishing.
\newblock \href {http://dx.doi.org/10.1088/0965-0393/20/8/085007} {\path{doi:10.1088/0965-0393/20/8/085007}}.
\newline\urlprefix\url{https://dx.doi.org/10.1088/0965-0393/20/8/085007}

\bibitem{lee_grain_2000}
S.~B. Lee, N.~M. Hwang, D.~Y. Yoon, M.~F. Henry, \href{https://doi.org/10.1007/s11661-000-1016-z}{Grain boundary faceting and abnormal grain growth in nickel}, Metallurgical and Materials Transactions A 31~(13) (2000) 985--994.
\newblock \href {http://dx.doi.org/10.1007/s11661-000-1016-z} {\path{doi:10.1007/s11661-000-1016-z}}.
\newline\urlprefix\url{https://doi.org/10.1007/s11661-000-1016-z}

\bibitem{zhao_abnormal_2015}
B.~B. Zhao, \href{https://doi.org/10.1179/1432891715Z.0000000001554}{Abnormal grain growth with 1 0 0 planar interface in the electrodeposited nickel}, Materials Research Innovations 19~(sup4) (2015) S251--S254, publisher: Taylor \& Francis \_eprint: https://doi.org/10.1179/1432891715Z.0000000001554.
\newblock \href {http://dx.doi.org/10.1179/1432891715Z.0000000001554} {\path{doi:10.1179/1432891715Z.0000000001554}}.
\newline\urlprefix\url{https://doi.org/10.1179/1432891715Z.0000000001554}

\bibitem{sal2022disconnection}
M.~Salvalaglio, D.~J. Srolovitz, J.~Han, {Disconnection-mediated migration of interfaces in microstructures: II. diffuse interface simulations}, Acta Materialia 227 (2022) 117463.
\newblock \href {http://dx.doi.org/10.1016/j.actamat.2021.117463} {\path{doi:10.1016/j.actamat.2021.117463}}.

\bibitem{kameswaran2006simultaneous}
S.~Kameswaran, L.~T. Biegler, Simultaneous dynamic optimization strategies: Recent advances and challenges, Computers \& Chemical Engineering 30~(10-12) (2006) 1560--1575.
\newblock \href {http://dx.doi.org/10.1016/j.compchemeng.2006.05.034} {\path{doi:10.1016/j.compchemeng.2006.05.034}}.


\end{thebibliography}
\end{document}